\documentclass[12pt]{article}
\pdfoutput=1 
\usepackage{amsmath}
\usepackage{amssymb}
\usepackage{enumerate}
\usepackage{natbib}
\usepackage{url} % not crucial - just used below for the URL 
\usepackage{amsfonts}
\usepackage{tikz}
\usepackage{graphicx}
\usepackage{mathtools}
\usepackage{enumitem}
\usepackage{bigints}
\usepackage{mathrsfs}
\usepackage{fancybox}
\usepackage{pdfpages}
\usetikzlibrary{arrows}
\usetikzlibrary{positioning}
\usepackage[colorlinks,citecolor=blue,urlcolor=blue]{hyperref}
\usepackage{appendix}
\usepackage{etoc}
\usepackage[english]{babel}

\usepackage{titlesec}

\titleformat*{\section}{\large\bfseries}

\usepackage[pagewise]{lineno}
\usepackage[mathscr]{eucal}
\usepackage{hyperref}
\usepackage{dlfltxbcodetips}
\usepackage{bbm} 
\usepackage{mathtools}
\usepackage{subfig}
\usepackage{amsfonts}
\usepackage{amsmath} 
\usepackage{amssymb}
\usepackage{fancybox} 
\usepackage{graphicx}
\usepackage{bm}
\usepackage{latexsym}

\usetikzlibrary{fit,positioning}

\newcommand{\condp}{\widehat{\tau}}

\newcommand{\mbE}{\mathbb{E}}

\newcommand{\mbP}{\mathbb{P}}

\newcommand{\mbR}{\mathbb{R}}

\newcommand{\mbX}{\mathbb{X}}
\newcommand{\mbY}{\mathbb{Y}}
\newcommand{\mbZ}{\mathbb{Z}}

\newcommand{\bx}{\bm{x}}
\newcommand{\by}{\bm{y}}

\newcommand{\bY}{\bm{Y}}

\newcommand{\balpha}{\mbox{\boldmath$\alpha$}}
\newcommand{\bphi}{\mbox{\boldmath$\phi$}}

\newcommand{\bmu}{\mbox{\boldmath$\mu$}}

\newcommand{\bSigma}{\mbox{\boldmath$\Sigma$}}

\newcommand{\dsum}{\displaystyle\sum\limits}

\newcommand{\dprod}{\displaystyle\prod\limits}

\newcommand{\btt}{\begin{box}}
\newcommand{\ett}{\end{box}}

\newcommand{\btheorem}{\begin{bclogo}[couleur={rgb:orange,0;yellow,0;white,1},arrondi=0.1,logo=\bcplume,ombre=true]{Theorem}}
\newcommand{\ettheorem}{\end{bclogo}}

\newcommand{\bst}{\begin{bclogo}[couleur={rgb:orange,1;yellow,1;white,0.5},arrondi=0.1,logo=\bcpanchant]}
\newcommand{\est}{\end{bclogo}}

\newcommand{\ba}{\begin{array}{llllllllll}}
\newcommand{\ea}{\end{array}}

\newcommand{\bea}{\begin{equation}\begin{array}{llllllllll}}
\newcommand{\eea}{\end{array}\end{equation}}

\newcommand{\be}{\begin{equation}\begin{array}{lllllllllllllllll}}
\newcommand{\beno}{\begin{equation}\begin{array}{lllllllllllll}\nonumber}
\newcommand{\ee}{\end{array}\end{equation}}

\newcommand{\bel}{\begin{equation}\begin{array}{lllllllllllll}\nonumber}
\newcommand{\eel}{\Box\end{array}\end{equation}}

\newcommand{\bi}{\begin{itemize}}
\newcommand{\ei}{\end{itemize}}

\newcommand{\ben}{\begin{enumerate}}
\newcommand{\een}{\end{enumerate}}

\newcommand{\bbq}{\begin{quote}\bf\em}
\newcommand{\ebq}{\end{quote}}

\renewcommand{\=}{&=&}

\newcommand{\hide}[1]{}
\newcommand{\ghost}[1]{}

\newcommand{\s}{\vspace{0.25cm}}

\newcounter{comment}
\newenvironment{comment}[1][]{\refstepcounter{comment}\par\smallskip\noindent%
\textbf{Comment~\thecomment #1}:\vspace{0.25cm}\\ \rmfamily}{}
\newcommand{\bcom}{\begin{comment}\em}
\newcommand{\ecom}{\end{comment}}

\newcounter{ex}

\newcounter{counterexample}

\newcounter{definition}

\newcounter{theorem}
\newenvironment{theorem}[1][]{\refstepcounter{theorem}\par\smallskip\indent%
\textbf{Theorem~\thetheorem #1}.\em \rmfamily}{}

\newcounter{proposition}

\newcounter{ttproof}
\newcounter{pproof}
\newcounter{corollary}

\newcounter{ccproof}
\newcounter{llproof}
\newcounter{lemma}

\newcounter{example}

\newcounter{com}

\newcounter{assumption}

\newcommand{\blind}{1}

\DeclareMathOperator{\Var}{\operatorname{Var}}

\begin{document}

\bibliographystyle{plainnat}

\def\spacingset#1{\renewcommand{\baselinestretch}%
{#1}\small\normalsize} \spacingset{1}

%%%%%%%%%%%%%%%%%%%%%%%%%%%%%%%%%%%%%%%%%%%%%%%%%%%%%%%%%%%%%%%%%%%%%%%%%%%%%%

\if1\blind
{
  \title{\bf Profile monitoring of random functions with Gaussian process basis expansions \vspace{.25in} 
  } 
  \author{ 
    Takayuki Iguchi \\
    Department of Mathematics \& Statistics, Air Force Institute of Technology
    \thanks{This research was supported by the Test Resource Management Center (TRMC) within the Office of the Secretary of Defense (OSD), contract \#FA807518D0002
    and the Environmental Security Technology Certification Program (EW24-7973).
    The views expressed are those of the authors and do not reflect the official views of the United States Air Force, nor the Department of Defense.
    Mention of trade names, commercial products, or organizations do not imply endorsement by the U.S. Government. 
    Imagery in this document are property of the U.S. Air Force. 
    }\\\\
    Jonathan R. Stewart\\
    Department of Statistics, Florida State University 
    \\\\
    Eric Chicken \\
    Department of Statistics, Florida State University \\\\
  }
  \date{June 20, 2025}
  \maketitle
} \fi

\if0\blind
{
  \bigskip
  \bigskip
  \bigskip
  \begin{center}
    {\LARGE\bf Profile monitoring of random functions with Gaussian process basis expansions} 
\end{center}
  \medskip
} \fi

\bigskip
\begin{abstract}
We consider the problem of online profile monitoring of random functions that admit basis expansions possessing random coefficients for the purpose of out-of-control state detection. Our approach is applicable to a broad class of random functions which feature two sources of variation: additive error and random fluctuations through random coefficients in the basis representation of functions. We focus on a two-phase monitoring problem with a first stage consisting of learning the in-control process and the second stage leveraging the learned process for out-of-control state detection. The foundations of our method are derived under the assumption that the coefficients in the basis expansion are Gaussian random variables, which facilitates the development of scalable and effective monitoring methodology for the observed processes that makes weak functional assumptions on the underlying process. We demonstrate the potential of our method through simulation studies that highlight some of the nuances that emerge in profile monitoring problems with random functions, and through an application. 
\end{abstract}

\noindent%
\vfill

\newpage
\spacingset{1.8} % DON'T change the spacing!

\section{Introduction} 
\label{sec:intro}

%---edit below
Determining if a process of interest is changing over time given observations of quality characteristics is the goal of statistical process control (SPC). 
A common tool for this monitoring task is the control chart. 
Control charts have many applications including, for example, in semiconductor manufacturing \citep{Gardner1997},
monitoring a stamping operation force \citep*{Jin1999}, 
artificial sweetener manufacturing \citep{Kang2000}, 
mechanical component manufacturing \citep*{Colosimo2008}, 
automobile engineering \citep*{amiri2009}, 
and public health surveillance \citep{Bersimis2022PublicHealthMSPC}. 
The design of a monitoring method is dependent on the assumptions on the stochastic nature of the process of interest. 
Given the increasing complexity of the processes of interest we wish to monitor, practitioners need a flexible approach which makes the least assumptions on the data we observe.  
We consider a quality characteristic to be represented by a functional relationship between a set of covariates and a response called a {\it profile}. 
% Monitoring profiles will require multiple observations of  
%---edit above
In the literature, control charts acting on profiles have commonly made either strong (parametric) assumptions on the form of the profile or use a large number of observations of a profile. 
Nearly all profile monitoring approaches require the functional relationship between the response and the covariate to be fixed. 

We relax this assumption by considering the problem of profile monitoring in functional situations where the function itself is a random object. 
To introduce our problem, consider infinite dimensional random processes $\{Y_x\}_{x \in \mbX}$ of the following form:   
\be
\label{eq:Y}
Y_x
\= f(x, \balpha)  + \epsilon_x, 
\ee
for $x \in \mbX \subset \mbR$ 
where $\balpha \in \mbR^k$ ($k \in \mbZ^{+}$) is a random vector with unknown distribution function, 
$\mbX \subset \mbR$ is a compact subset of $\mbR$ (which will typically be a closed and bounded interval, 
e.g., $\mbX = [a, b]$),  
and $\{\epsilon_x\}_{x \in \mbX}$ is a collection of jointly independent random variables
which are normally distributed with mean zero, 
i.e.,
$\epsilon_x \sim \text{Norm}(0, \sigma_{\epsilon,x}^2)$. 
We assume that the function $f(x, \balpha)$ 
admits a basis expansion with respect to a given set of $K$ basis functions 
$\phi_{k} : \mbX \mapsto \mbR$ ($k = 1, \ldots, K$) 
collected into the vector 
$\bphi(x) = (\phi_1(x), \ldots, \phi_K(x))$ 
with the following form 
\be
\label{eq:basis}
f(x, \balpha) 
\= \langle \balpha, \, \bphi(x) \rangle 
\= \dsum_{k=1}^{K} \, \alpha_k \, \phi_k(x),
&& x \in \mbX. 
\ee
Observe that in our problem setup, 
the function $f(x, \balpha)$ is a random function, 
deviating from other approaches taken in the literature
which assume constant in-control functions with additive noise in the form of independent mean zero Gaussian perturbations. 
% ***
Thus, our in-control characterization can be thought of as a class of functions, rather than a single function.
% ***
As a result, 
there are two sources of randomness and variation in our setup: 
\ben
\item {\bf Sampling error:} The realization of the process $\{Y_x\}_{x \in \mbX}$  
depends on an underlying random vector $\balpha$,
which we may regard as a random effect. 
The value of $\balpha$ will determine the baseline function $f$ per \eqref{eq:Y} and \eqref{eq:basis}. 
\item {\bf Measurement error:} We treat the signature of random variables 
$\{\epsilon_x\}_{x \in \mbX}$ as independent measurement error 
associated with obtaining a noisy observation of the function $f(x, \balpha)$ 
through the observation process $\{Y_x\}_{x \in \mbX}$.
While we do not assume that the error terms must be identically distributed, 
we do assume that measurement error is random, 
additive,
mean zero,  
and independent across indices of the process.   
\een
An important aspect of this class of random functions which motivates our monitoring methodology is that the random variables $f(x_1, \balpha)$ and 
   $f(x_2, \balpha)$ will---in general---be dependent for $x_1 \neq x_2$,
a point which we will emphasize through illustrative examples later.

It is important to note that our interest in the basis representation in \eqref{eq:basis} is not to provide a means to estimate functions, 
but rather to establish the statistical foundations of a class of random functions 
for motivating the development of our monitoring methodology.
% ***
In fact, our proposed methodology does not require estimation of either the vector $\balpha$ or the basis functions $\{\phi_k\}_{k=1}^K$. 
% ***
A convenient property of the basis representation in \eqref{eq:basis} is that the functional dependence of 
$f(x, \balpha)$ is linear in the random coefficients $\balpha$. 
Under the assumption that $\balpha \sim \text{MvtNorm}(\bmu_{\balpha}, \bSigma_{\balpha})$ 
for some mean vector $\bmu_{\balpha} \in \mbR^k$ 
and $(K\times K)$-dimensional positive definite covariance matrix $\bSigma_{\balpha}$.
As a result of this assumption and the form of \eqref{eq:basis}, 
the functional values $f(x, \balpha)$ will be Gaussian random variables. 
We call functions $f(x, \balpha)$ admitting the form of \eqref{eq:basis} with the Gaussianity assumption 
{\it random functions with Gaussian process basis expansions}.

Through the representation as a random function with a Gaussian process basis expansion, 
the realized observation process $\{Y_x\}_{x \in \mbX}$ is itself a Gaussian process,  
noting that for any discretely realized set of observation times 
$x_1 < x_2 < \ldots < x_n$ ($n \in \mbZ^{+}$) in $\mbX$, 
$(Y_{x_1}, \ldots, Y_{x_n}) \sim \text{MvtNorm}(\bmu, \bSigma)$ 
for some mean vector $\bmu \in \mbR^m$ and covariance matrix $\bSigma \in \mbR^{m \times m}$.
We will leverage this distributional consequence in the coming sections 
in order to establish effective and tractable learning and monitoring methodology.

While the observation process $\{Y_x\}_{x \in \mbX}$ is infinite dimensional, 
any realized observation of the infinite dimensional process will be finite dimensional.
Concretely, 
we consider a set of $n \in \mbZ^{+}$ pre-chosen {\it monitors} $\bx = (x_1, \ldots, x_n)$ 
($x_i \in \mbX$) at which we observe the process.  
Hence, 
an observation of the process $\{Y_x\}_{x \in \mbX}$ 
at monitors $\bx$ 
results in an vector of observations $\by_{\bx} = (y_{x_1}, \ldots, y_{x_m})$,
where each $y_{x_i}$ is an observation of the component $Y_{x_i} = f(x_i, \balpha) + \epsilon_{x_i}$ at monitor $x_i \in \mbX$. 
Distinct from current approaches in the literature, 
our set-up does not preclude the possibility that the observations 
$(y_{x_1}, \ldots, y_{x_n})$ will be dependent. 
While the additive error terms $(\epsilon_{x_1}, \ldots, \epsilon_{x_n})$ are assumed to be independent random variables, 
the functional observations $(f(x_1, \balpha), \ldots, f(x_n, \balpha))$ 
may be dependent (and generally will be dependent) depending on the specification of $f$, 
as the function value at each point $x_1, \ldots, x_n$ are allowed to depend on the value of the random vector $\balpha$
through the form in \eqref{eq:basis}.  
We make this point clear in the following example. 

Finally,
we point out that our interest in a class of random functions with Gaussian process basis expansions is to provide concrete statistical foundations for our profile monitoring problem,
that is to develop a class of random functions $f(x, \balpha)$ 
for which the functional response $Y_{\bx}$ will be Gaussian. 
This setup does not deviate significantly from other approaches taken in the literature, 
which typically involve non-random functions with additive Gaussian noise,
which will produce Gaussian functional responses. 
As will be seen, 
our methodology operates under the essential assumption that the functional response is Gaussian, 
and notably is agnostic to the functional assumptions put in place to generate the distribution of the response. 
As such, 
we reiterate that our methodology would extend to any profile monitoring problem with Gaussian functional response. 

\s\s

\begin{figure}[t]
\centering 
\includegraphics[width = .95 \linewidth, keepaspectratio]{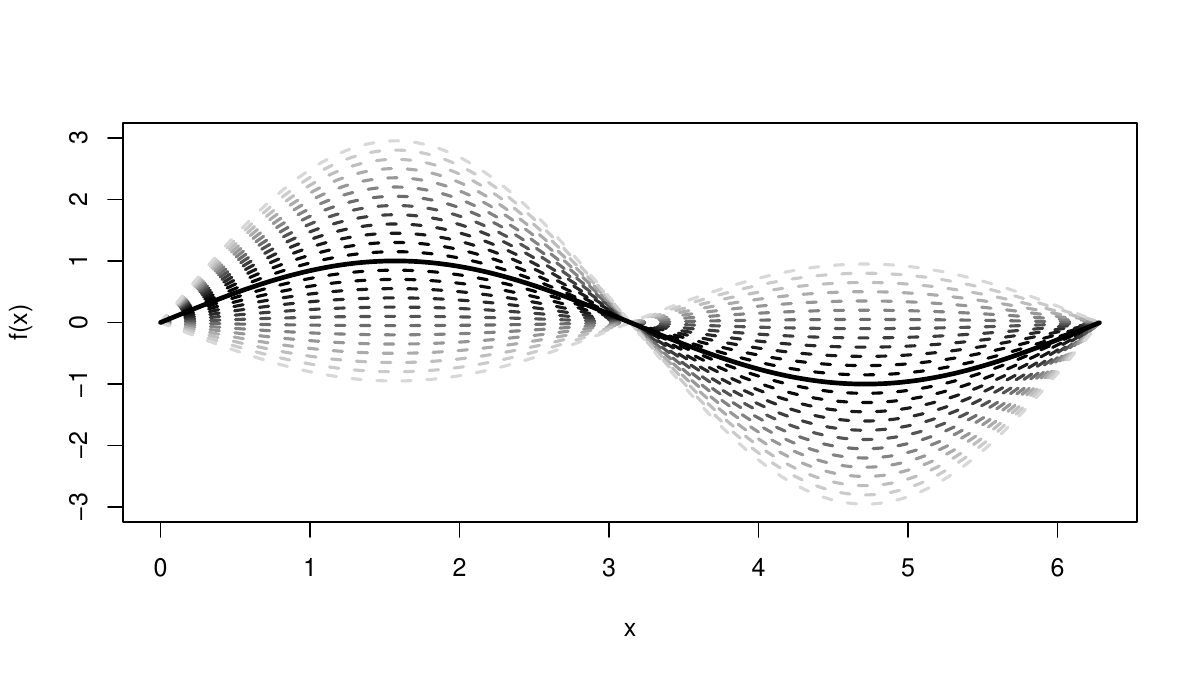}
\caption{\label{fig:sin} Visualization of $f(x) = c \, \sin(x)$ 
varying $c \in [-1, 3]$. The color of each curve corresponds to the likelihood of 
the $\text{N}(1,1)$ distribution evaluated at $c$,
where darker shades correspond to higher likelihoods and lighter shades to the lower likelihoods.} 
\end{figure}

\subsection{An illustrative example} 
\label{sec:example}

\begin{figure}[t]
\centering
\includegraphics[width = .32 \linewidth, keepaspectratio]{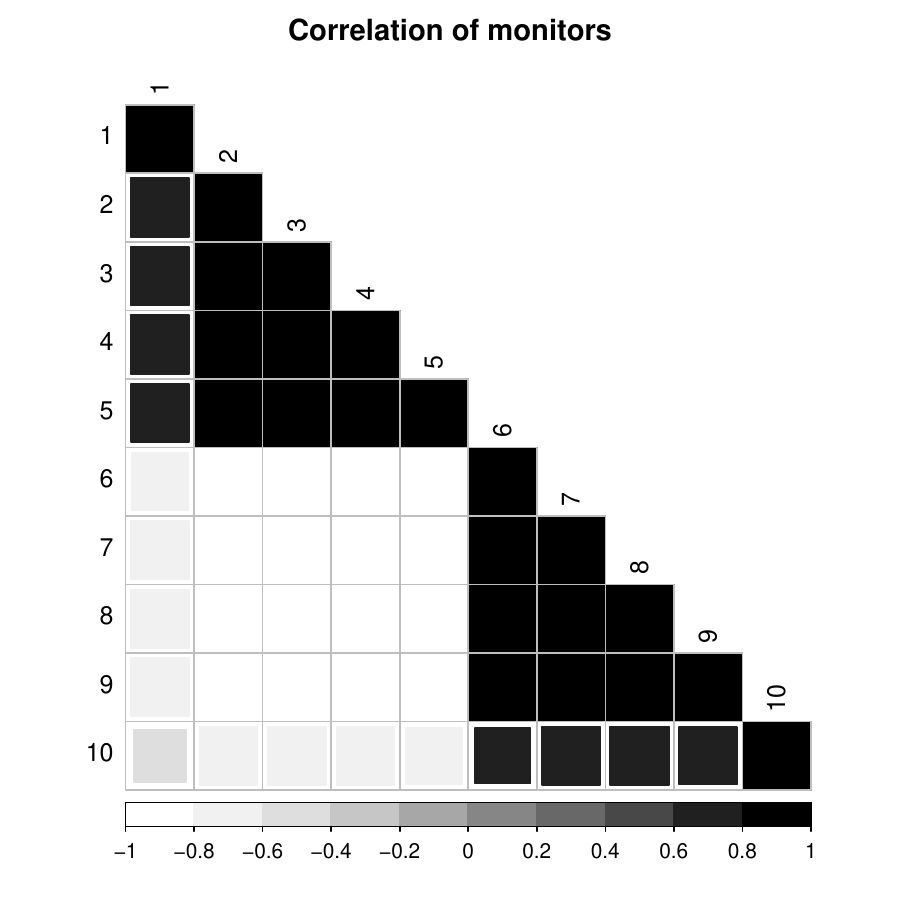} %
\includegraphics[width = .32 \linewidth, keepaspectratio]{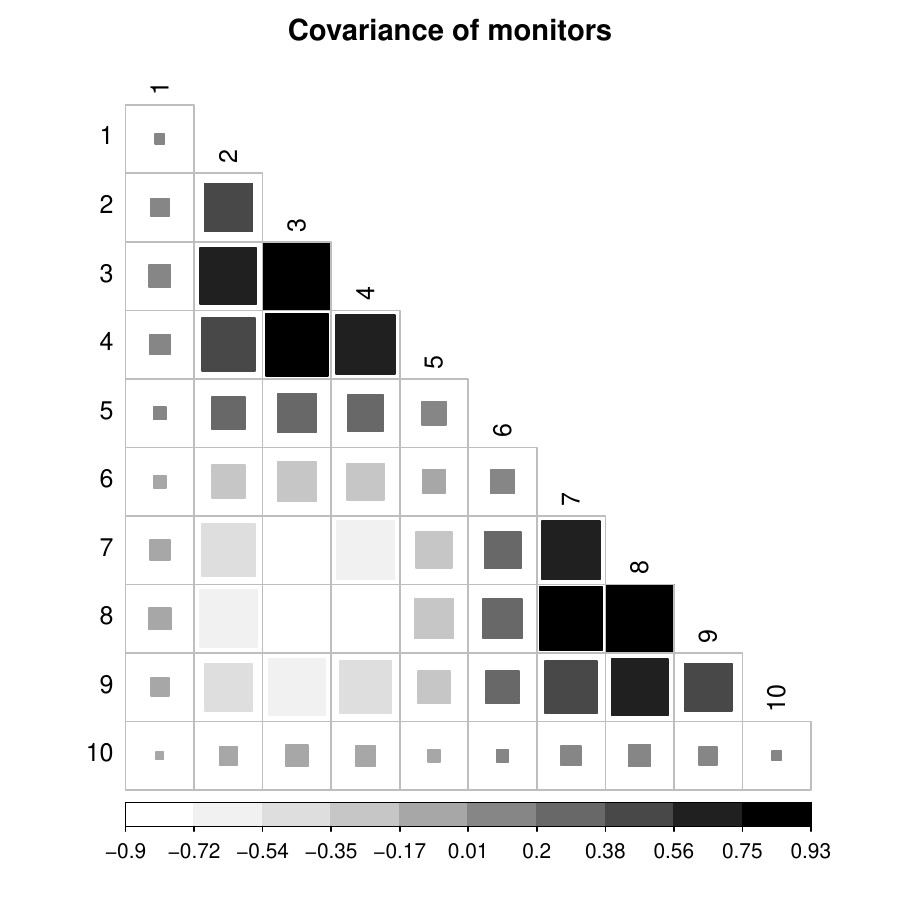} % 
\includegraphics[width = .32 \linewidth, keepaspectratio]{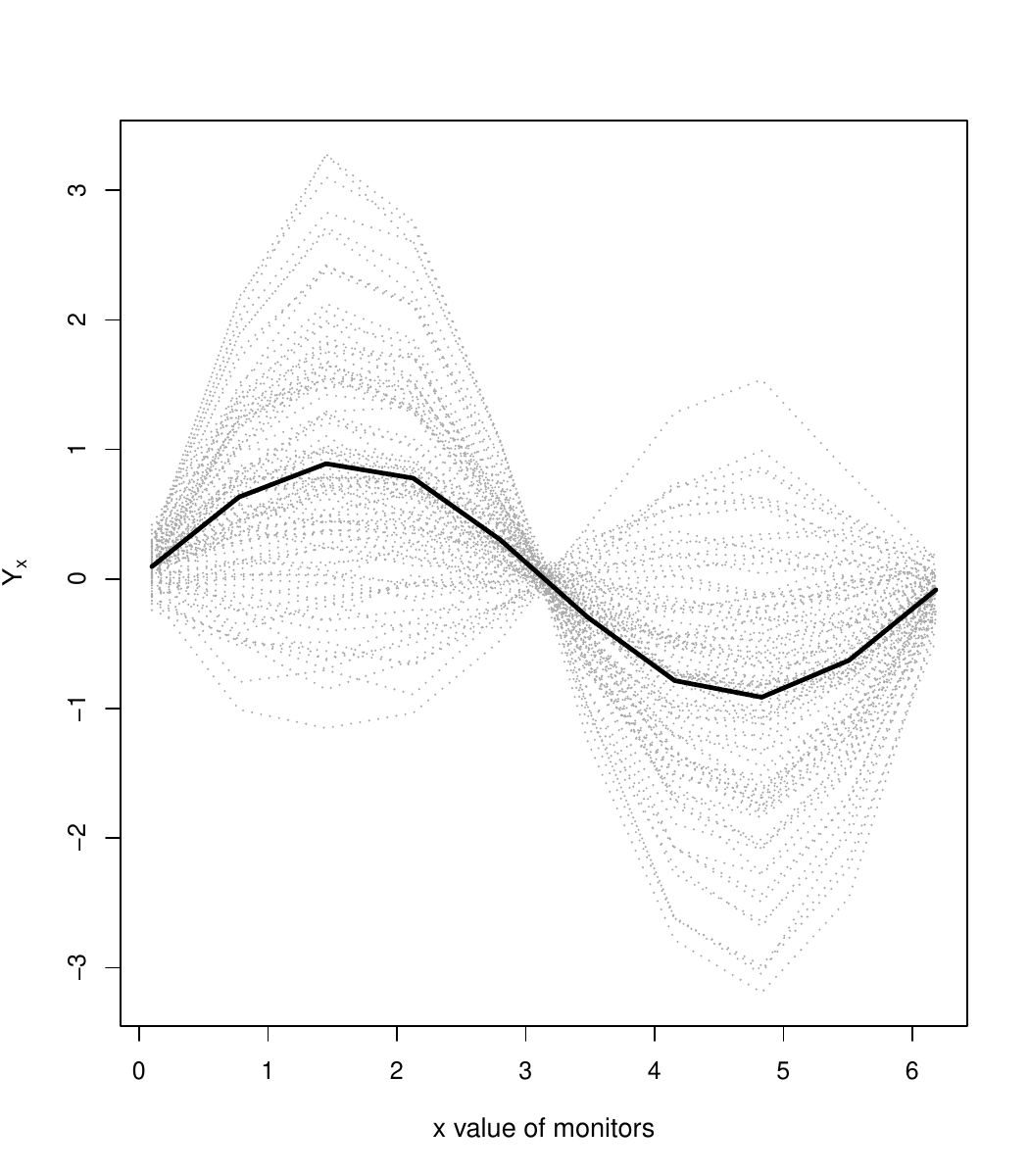}
\caption{\label{fig:sin_ex} Realization of $100$ replicates of the process $\bY_{\bx}$ 
at $10$ monitors $\bx = (x_1, \ldots, x_{10})$ equally spaced to cover the interval $[0, \, 2 \pi]$. 
Empirical correlations and covariances of the monitor sites are visualized with color 
corresponding to sign and magnitude and area of boxes corresponding to absolute magnitude. 
The $100$ replicates of $\bY_{\bx}$ are visualized in the third panel with the solid line representing 
the average value of the monitor sites.} 
\end{figure}

Let $f(x, \alpha) = \alpha \sin(x)$,
taking $\mbX = [0, \, 2 \pi]$,  
and assume $\alpha \sim N(1, 1)$. 
We visualize $f(x, \alpha)$ on the interval $x \in[0, \, 2 \pi]$ in Figure \ref{fig:sin}. 
Note that the these visualizations are of the random function $f(x, \alpha)$ 
and thus do not include the measurement errors of the random noise terms $\{\epsilon_x\}_{x \in \mbX}$.  
We visualize observations of the process $\{Y_x\}_{x \in \mbX}$ based on the function 
$f(x, \alpha) = \alpha \sin(x)$ and observed at $10$ monitor sites 
$\bx = (x_1, \ldots, x_{10})$ 
that are equally spaced to cover the interval $[0, \, 2 \pi]$,
taking $\epsilon_{x_i} \sim \text{N}(0, 0.1^2)$ ($i \in \{1, \ldots, 10\}$), 
as well as the correlations and covariances between these $10$ monitor sites,
in Figure \ref{fig:sin_ex}.

An important consequence of this modeling framework lies in the fact that,
despite the measurement errors $(\epsilon_{x_1}, \ldots, \epsilon_{x_n})$ being independent, 
there is high correlation between the process values $\bY_{\bx} = (Y_{x_1}, \ldots, Y_{x_n})$
at the various monitor sites $\bx = (x_1, \ldots, x_n)$.  
This is due to the fact that the values of the random function $f(x, \alpha)$ 
at different values of $x \in \mbX$ will depend on the same underlying random variable $\alpha$,
i.e.,
$f(\pi \,/\, 2) = \alpha$ and $f(3 \, \pi \,/\, 2) = - \alpha$. 
In this specific example, 
the value of $\alpha$ controls the amplitude of the sine function,
which can be seen in Figure \ref{fig:sin}. 
The component-wise variation in the process can be large,
as the deviations shown in the third panel of Figure \ref{fig:sin_ex} 
demonstrate the potential for the function $f(x, \alpha)$ and therefore the observation process
$\bY_{\bx} = (Y_{x_1}, \ldots, Y_{x_n})$
to exhibit large deviations from the mean process values $\{\mu_x\}_{x \in \mbX}$ 
with the definition $\mu_x \coloneqq \mbE \, Y_x$ ($x \in \mbX$),
visualized with a solid line in Figure \ref{fig:sin}.
Ordinarily,
such large deviations could lead to very low signal-to-noise ratios, 
resulting in a challenging monitoring problem in certain circumstances. 
In spite of this, 
we are able to exploit the correlation structure of monitor sites evidenced in Figure \ref{fig:sin_ex} 
in order to develop effective statistical monitoring methodology for the purpose of out-of-control state detection.

\subsection{Background}\label{sec:background}

% DETAIL PROFILE MONITORING PROBLEM SPECIFICALLY

Profile monitoring is performed is two phases. 
In Phase I, the aim is to define an in-control process and to identify from a finite set of observations which ones are in-control.  
Using knowledge gained from Phase I via either a parametric form of an in-control process or a set of $m$ observations deemed to be in-control, a Phase II analysis aims to detect a deviation from an in-control process using  $\{(y_{x_1}^{(j)},\dots, y_{x_n}^{(j)}): j=-m+1,\cdots,0,\cdots,t\}$ of quality characteristics at each discrete time $t\geq 1$.
At each time $t$, a control chart can claim a process is in-control or out-of-control,
that latter of which is defined as anything not in-control.
We consider a change-point framework wherein there is some time $\tau$ which denotes the last in-control observation (and the first out-of-control observation is made at time $\tau + 1$). 
As a result, 
there are two errors that can be made in this setting. 
A false alarm is raised if the control chart claims the process is out-of-control at some time $t \leq \tau$ when the process is truly in control. 
A measure of this kind of error is the False Alarm Rate (FAR) which is the probability of the next alarm being a false alarm wherein the control chart is reset after each false alarm. 
The other type of error is to incorrectly claim a process to be in-control. 
A common measure of this error is the out-of-control average run length ($ARL_1$), which is defined as the expected number of out-of-control profiles observed until a true alarm is raised. 
Typically, an alarm is raised at time $t$ if the monitoring statistic using $\{\bm{y}^{(j)}_{\bm{x}}\}_{j=-m+1}^t$ lies outside of an interval of lower and upper control limits. 
Similar to traditional hypothesis testing where the rejection region is established prior to conducting the test to achieve a desired level of significance, the control limits are calibrated to ensure a desired in-control average run length ($\text{ARL}_0$) prior to starting any Phase II  monitoring.

% CATEGORIZE PROFILE MONITORING METHODS
Profile monitoring methods typically make assumptions on the functional relationship $f$ between the covariates and the response and the additive error $\epsilon_x$.
These methods can be categorized by these functional and stochastic assumptions in addition to their approach to estimating $f$. 
Poor performance can occur if these assumptions are misaligned with the true nature of the random process. 
For example, if the wrong parametric assumptions on $f$ are made, the performance of such a control chart would suffer. 
Moreover, it is often the case where a practitioner has access only to $m$ historical profiles (observed at times $1-m, \dots, 0$ under an in-control process), so exact parametric forms of an in-control functional relationship may not be available. 
This motivates research in nonparametric profile monitoring methods. 
The nonparametric approaches to profile monitoring thus far have required a large number of monitors $n$ on the order of 100. 
With such a large sample size and a finite space in which to place these monitors, the location of the monitors tend to be close together. 
As such, the measurements between monitors tend to be correlated. 
This motivates techniques which account for within-profile correlation.

Table \ref{tab:within-profile correlation profmon comparison} shows a selection of profile monitoring methods which account for within-profile correlation. 
Notice methods which make parametric assumptions require much fewer monitors in a profile than their nonparametric counterparts. 
A key contribution of this work is that we provide a nonparametric approach which accounts for within-profile correlation which works well for small $n$. 

We contrast our method with a parametric method and a nonparametric method.
In the two categories, the two methods closest to our monitoring scheme in approach are those of the principal component based approach of \cite{Colosimo2010Comparison} and the Tweedie exponential dispersion process approach of \cite{li2023TweedieProfmon}.   

\cite{Colosimo2010Comparison} performs PCA on the responses of a profile with fixed monitoring points. 
A Hotelling $T^2$ control chart is then employed on the retained principal components. 
Their approach requires large sample sizes to achieve power comparable to other methods in the literature. 
The work of \cite{li2023TweedieProfmon}, although parametric in nature, attempts to use a parametric form which can encompass a large variety of profiles by modeling the within-profile increments ($Y_{s_i} - Y_{s_{i-1}}$).
In so doing, they assume the increments are independent 
and require the responses to be non-negative. 
Although our approach places a Gaussian assumption on the errors and responses, our approach does not require the independent increment assumption and does not place any specific assumption on the functional form of the profiles. 
We introduce and discuss our proposed approach in the next section.

\begin{table}[]
    \centering
\scalebox{0.8}{
    \begin{tabular}{p{5cm}|p{5cm}|p{6cm}|c}
    Citation
    & Profile type
    & Assumptions
    & Sample size\\ \hline \hline
    \citep{Colosimo2010Comparison}
    & Nonparametric profile (roundness profiles)
    & fixed sensor locations
    & 748\\ \hline
    \citep{Chang2010}
    & Nonparametric Nonlinear Profile
    & Equispaced monitors, Gaussian errors 
    & 240\\ \hline
    \citep{cheng2018linearprofilewithinARMA}
    & Linear profile 
    & Within profile errors have ARMA structure
    & 150, 300\\ \hline
    \citep{fan2019NMLE_AR1errors_T2}
    & Polynomial \& Nonlinear mixed effects model
    & Errors follow an AR(1) process
    & 48\\ \hline
    \citep{xia2019LinearProfileWaldTypeStatistic} 
    & Linear profile 
    & i.i.d. additive noise
    & 10 \\ \hline
    \citep{ding2023GPregressionprofile}
    & Simple profile with heterogeneous errors
    & Gaussian kernel used and sensor locations must differ over time
    & 20\\ \hline
    \citep{song2022lGLM_randompred_parawincorr}
    & Generalized Linear Model
    & Random predictors and a known parametric within profile correlation structure 
    & 10 \\ \hline
    \citep{siddiqui2019LMM_Pspline_onResiduals,nassar2021LMM_MEWMA, nassar2022LMMProfileViaResiduals}
    & Linear mixed model
    & Pure error defines within profile correlation structure
    & 10, 20\\ \hline
    \citep{li2023TweedieProfmon}
    & Parametric
    & Responses follow a Tweedie exponential dispersion process
    & 5, 10
    \\ \hline
    This paper
    & Random nonparametric profile
    & Functional relationship is a random function with a Gaussian process basis expansion
    & 10
    \end{tabular}
}
    \caption{A comparison of recent profile monitoring papers allowing for within-profile correlation. 
    Notice the competing literature either make strong assumptions on the profile type or leverage a large sample size. 
    }
    \label{tab:within-profile correlation profmon comparison}
\end{table}

%--- edit above

\section{Monitoring methodology} 
\label{sec:monitor}

We consider monitoring the observation process $\{Y_x\}_{x \in \mbX}$,
specified via \eqref{eq:Y} and observed at $n \in \mbZ^{+}$ distinct and ordered (from smallest to largest) monitor sites
$\bx = (x_1, \ldots, x_n)$, 
for the purpose of out-of-control state detection. 
Following standard practices in the statistical process control literature, 
we break the monitoring problem down into two parts:
\ben
\item {\bf Control limit calibration:} Learn the in-control process from a reference data set of sample size $m \in \mbZ^{+}$  
which is a random sample $\by^{(-m+1)}_{\bx}, \ldots, \by^{(0)}_{\bx}$ of observations 
known to have derived from the in-control process. 
From these reference data, we use bootstrapping techniques to obtain a control limit achieving a desired $\text{ARL}_0$.

\item {\bf Online monitoring:} Monitor streams of new observations $\by_{\bx}^{(t)}$ ($t \in \{1, 2, \ldots\}$) 
from unknown processes (i.e., either from the in-control process or another process), 
under the assumption that $\by_{\bx}^{(t)}$ ($t\leq \tau$) 
are observations of the in-control process and $\by_{\bx}^{(t)}$ ($t > \tau$) 
are observation of a process distinct from the in-control process,
which we will refer to as the {\it out-of-control} state. 
\een
The statistical problem can be divided into two corresponding problems for each of the two parts described above.
The first is an estimation problem, 
where we want to estimate a statistical model for the in-control process based on a random sample 
assumed to be sampled from the in-control process.
Given the estimated model of the random in-control process, we estimate a control limit which will provide us a desired in-control run length $ARL_0$. 
The online monitoring problem is an inference problem, where our fundamental goal is to minimize the distance between two times: time $\widehat{\tau}$
(the time at which we stop the monitoring process) and time $\tau$ the time at which the process goes out-of-control.

\subsection{Monitoring statistics}\label{ssec: monitoring statistics}

We consider the following statistics 
for constructing our monitoring statistic: 
\beno
p_{j,i}
&\coloneqq& \min\left\{ \,\,
\mbP(Y_{x_j} < y_{x_j}^{(i)} \;|\; Y_{x_l} = y^{(i)}_{x_l}, 1 \leq l \neq j \leq n), \; \right. \\
&& \left. 
\hspace{.5in} 
\mbP(Y_{x_j} > y_{x_j}^{(i)} \;|\; Y_{x_l} = y^{(i)}_{x_l}, 1 \leq l \neq j \leq n)
\,\,\right\},
\ee
 which are essentially two-tail conditional $p$-values
 for each of the monitoring sites,
i.e.,
we compute the probability using the conditional distribution of $Y_{x_j}$ conditional on 
$Y_{x_l} = y^{(i)}_{x_l}$ ($1 \leq l \neq j \leq n$).
We can view the values $p_{1,i}, \ldots, p_{n,i}$ 
for iteration $i$ of the monitoring process as a conditional $p$-value which provides a measure of how consistent the observed process is at each monitor with regards to the rest of the observed process and the learned in-control process. 

An important question is how to utilize these conditional $p$-values $p_{j,i}$ in the online monitoring process. 
The multiple conditional $p$-values must be aggregated to obtain a single decision,
that of whether the process is in-control or out-of-control. 
To this end, 
we outline two methods for aggregating the conditional $p$-values:
\ben
\item {\bf Minimum rule:}
\beno
p_{\min,i}
&\coloneqq& \min(p_{1,i}, \ldots, p_{n,i})
\ee
\item {\bf Geometric mean rule:}
\beno
p_{\textrm{geo},i}
&\coloneqq& \sqrt[n]{\dprod_{j=1}^{n} \, p_{j,i}}
\ee
\een
The minimum rule will be most sensitive to local deviations from the expected in-control behavior. 
At times, 
this may be too sensitive to false alarms and a compromise which takes into account some averaging of the conditional $p$-values $p_{1,i},\ldots,p_{n,i}$ will be more robust to false alarms. 
Here, 
we suggest the geometric mean as one such alternative. 
Given a rule $\widetilde{p}_i$ ($i \in \{1, 2, \ldots\}$), 
such as one outline above, 
we are able to define our stopping criterion: 
\beno
\condp 
&\coloneqq& \inf\left\{ i \in \mbZ^{+} \,:\, 
\widetilde{p}_i < r
\right\},
\ee
where here $r \in (0, 1)$ should be selected in order to appropriately calibrate the monitoring process,
typically calibrating to a desired average run length under the in-control process,
i.e.,
calibrating to a desired value of $\text{ARL}_0$. 

It is important to note that the conditional probabilities  
$p_{j,i}$ defined above
monitor both 
deviations in the values of the process $Y_{x}$ from the mean process $\mu_x$, 
as well as changes in the correlation structure.
The conditional formulation is motivated 
from the observation that the process definition in \eqref{eq:Y} and \eqref{eq:basis}
can give rise to strong correlations among observed process values at the monitor sites,
due to the functional dependence induced at monitor sites via $f(x, \balpha)$.  
We exhibited this in the example from the previous section in Figure \ref{fig:sin_ex}.  
The Gaussianity of $\bY_{\bx}$ implies that the conditional probabilities will have a tractable form as
if $\bY^{(i)} \sim \text{MvtNorm}(\bm{\mu}, \bm{\Sigma})$,
then
\beno
Y_{x_j} \,|\, \left(Y_{x_l} = y^{(i)}_{x_l}, \; 1 \leq l \neq i \leq n\right)
&\sim& \text{MvtNorm}\left(
\mathcal{U}_i(\by_{\bx}^{(i)}), \; 
\mathcal{V}_i(\by_{\bx}^{(i)}) \right),
\ee
where
\beno
\mathcal{U}_j(\by_{\bx}^{(i)})
\= \mu_j + \bm{\Sigma}_{-j,j} \, \bm{\Sigma}_{-j,-j}^{-1} \, (\by_{\bx_{-j}}^{(i)} - \bm{\mu}_{-j}) \\
\mathcal{V}_j(\by_{\bx}^{(i)})
\= \Sigma_{j,j} - \bm{\Sigma}_{j,-j} \, \bm{\Sigma}_{-j, -j} \, \bm{\Sigma}_{-j, j},
\ee
where we understand the indexing $\bm{\Sigma}_{-j,j}$ to be the column vector consisting
of the $j$th column of $\bm{\Sigma}$ excluding the $j$th row,
$\bm{\Sigma}_{j,-j}$ to be the equivalent for row-vectors,
$\bm{\Sigma}_{-j,-j}$ to be the $(m-1)\times(m-1)$ sub-matrix of $\bm{\Sigma}$
that excludes the $j$th row and column,
and
$\by_{\bx_{-j}}^{(i)}$ and $\bm{\mu}_{-j}$ to be the respective $(m-1)$-dimensional column vectors which
exclude the $j$th entry.
Using the above forms, 
the conditional probabilities $p_{j,i}$ are computationally tractable,
which makes computing monitoring statistics for the online 
monitoring problem feasible. 
A final point lies in the fact that both the mean vector $\bmu_{\bx} = \mbE \, \mbY_{\bx}$ 
and the variance-covariance matrix $\bSigma_{\bx} = \text{Var} \, \bY_{\bx}$ need to be learned from data. 
When the number of monitoring sites $n$ is sufficiently small relative to the size $m$ of the historical data set
$\bY_{-m}, \ldots, \bY_{-1}$ known to be in-control,  
the sample average and variance-covariance matrices computed off of the historical data set 
will provide both consistent (and asymptotically efficient) estimates of these quantities:
\beno
\widehat\bmu
\= \dfrac{1}{m} \, \dsum_{i=1}^{m} \, \bY_{-i} \s\\
\widehat\bSigma 
\= \dfrac{1}{m-1} \, \dsum_{i=1}^{m} \, (\bY_{-i} - \widehat\bmu) \, (\bY_{-i} - \widehat\bmu)^{\top}. 
\ee 
We will discuss extensions and refinements to our methodology when this assumption may not be satisfied later. 

A common approach in the statistical process control literature is to 
calibrate the stopping rule against the reference data set of in-control process observations 
to design decision rules that will give rise to an acceptable False Alarm probability,  
and aim to design a Phase I and II methodology that accurately learns the in-control process 
and possess good statistical power for detecting deviations from the in-control process state. 
We assume that each monitoring rule has been appropriately calibrated. 

\subsection{Calibrating control limits}

We have established the statistics we wish to monitor, and we now need to determine our lower control limit. 
Our approach which we cover in the following sections is two-fold:  (1) modify an existing univariate nonparametric control chart from \cite{willemain1996CC_EmpiricalRefDist} and (2) use semiparametric bootstrap to obtain a sufficiently large sample size to adopt the procedure in (1) for our profile monitoring purposes.

\subsubsection{Control limits on a control chart using order statistics}

In this section we describe a tuning-parameter-free modification to the nonparametric control chart developed in \cite{willemain1996CC_EmpiricalRefDist} and present new theoretical results on the control chart.

\cite{willemain1996CC_EmpiricalRefDist} originally derived the use of an empirical reference distribution to set control limits. 
They assume $U_1, \dots, U_m$ are a random sample of known in-control data.
For now, we the reader may consider $U_1, \dots, U_m$ to be $m$ monitoring statistics drawn from an in-control process prior to monitoring beginning. 
We denote the order statistics of $U_1, \dots, U_m$ by $U_{(1)}<\cdots< U_{(m)}$ and denote by $u_{(1)} < \cdots < u_{(m)}$ the observed values of these order statistics.
We call the intervals $(U_{(j-1)}, U_{(j)})$ (for $j \in\{1, \dots, m\} $) {\it blocks}, where $u_{(0)} = -\infty $ and $u_{(m+1)} = \infty$ by convention.
Now consider a sequence $V_1, V_2, \dots$ which are independent and identically distributed with $U_1, \dots U_m$.
The reader may treat these as monitoring statistics obtained in an online fashion during Phase II monitoring. 
Given a choice of $j$ and $b$ and observed values $u_{(1)}, \dots, u_{(m)}$, 
we define the in-control run length of their control chart to be $R_{wr} = \min\left\{ t \,:\, V_t \not\in (u_{(j)}, \, u_{(b+j)})\right\}$. 
It is worth noting that $R_{wr}$ defines a sequence of random variables indexed by $m$,
and that all of the following limiting results for $R_{wr}$ will be taken with respect to increase sequence in $m$. 
By way of iterated expectations, \cite{willemain1996CC_EmpiricalRefDist} show 
\beno 
\text{ARL}_0 \= 
\mbE_P[\mbE[R_{wr} \, | \, P]] 
\= \dfrac{m}{m-b}, 
\ee
where $P$ is the probability of $V_1$ falling in the block $(U_{(j)}, U_{(b+j)})$.
They separately compute 
\beno 
\Var[R_{wr}] 
\= \left(\dfrac{(b/m)}{(1 - (b/m))^2}\right) \, 
\left( \dfrac{ 1 - (b/m)  + 1/m}{ 1- (b/m) - 1/m}\right),
\ee
which shares similarities in structure with the variance of the Geometric distribution.
Assuming $b$ is chosen as a function of $m$ so that $\lim_{m\to\infty} b/m = p^\prime$ for some $p^\prime \in [0,1]$, 
then the first factor is the variance of the Geometric distribution and the second factor approaches $1$ in the limit as $m \to \infty$.
\cite{willemain1996CC_EmpiricalRefDist} do not derive the asymptotic distribution of these run lengths, 
which is a contribution of this work.

\cite{arts2004NonparametricPredictiveCC} provides a slightly more general version of the same control chart, but both of these approaches require at least one choice by the practitioner even after selecting a particular target $\text{ARL}_0$ for the control chart.  
\cite{willemain1996CC_EmpiricalRefDist} requires a choice of $j$ as $b$ is determined by the choice of $\text{ARL}_0$. 
\cite{arts2004NonparametricPredictiveCC} requires a choice of the $b$ blocks to form the acceptance region due to the flexibility of their approach.

The following theorem with proof in the supplementary materials provides the $\text{ARL}_0$ without the use of blocks. 

\s

\begin{theorem}\label{thm: main}
Let $U_1, \dots, U_m$ and $V_1, V_2, \dots$ be independent and identically distributed continuous random variables (i.e., monitoring statistics).
If $W = \min \{t \in \mathbb{Z}^+ : V_t < U_{(k)}\}$ and $2 \leq k < m$,
then $\mbE[W] = \frac{m}{k-1}$.
\end{theorem}

\s

We note that this result is a special case of results established by \citet[p. 32]{willemain1996CC_EmpiricalRefDist} and \citet[p. 11]{arts2004NonparametricPredictiveCC}. 
Specifically, letting $j=0$ and $b = m-k+1$ turns the control chart from \cite{willemain1996CC_EmpiricalRefDist} into our control chart. 
Setting the triplet $(b, m, n)$ in the notation of \citet{arts2004NonparametricPredictiveCC} to $(m-k+1,\, 1,\,  m)$ in our notation similarly shows our control chart is a special case of their more general control chart.
In contrast to the cited works, we prove Theorem \ref{thm: main} without using a notion of `blocks' or purely combinatorial arguments as was done in \cite{arts2004NonparametricPredictiveCC}.
Our approach also yields insights as to why using $U_{(1)}$ as a lower control limit is inadvisable and the restriction on $k\geq 2$ is necessary beyond ensuring that the quantity $\frac{m}{k-1}$ is well defined. 
If $U_{(1)}$ is used as a control limit, a series listed in Lemma 4 in the supplementary materials, which is used in the proof for Theorem \ref{thm: main}, diverges implying $\text{ARL}_0 = \infty$. 
\cite{willemain1996CC_EmpiricalRefDist} discourages using extreme order statistics as control limits, but a mathematical justification for poor behavior when $k=1$ is new to the literature.
\cite{willemain1996CC_EmpiricalRefDist} demonstrated the first two moments of the run length from our control chart asymptotically match that of a Geometric random variable. 
In contrast, we provide a stronger distributional result detailed below with a corresponding proof in the supplementary materials that establishes the Geometric distribution as the limiting distribution.

\s

\begin{theorem}\label{thm: convergegeo}
Consider the setting of Theorem \ref{thm: main}. 
Let $R_m = \min \{t \in \mathbb{Z}^+ : Y_t < X_{(k^*)}\}$
with $k^* = 1 + \frac{m}{\text{ARL}_0}$. 
Then $R_m \overset{D}{\to} G$ as $m \to \infty$, 
where $G \sim \operatorname{Geometric}\left(\frac{1}{\text{ARL}_0}\right)$.
\end{theorem}

\s

As a simulation study in Section \ref{sec:simulations} shows, $m$ must be fairly large for $R$ to be treated as a geometric variable. 
Many applications of statistical process control applications do not have a sufficiently large $m$ for Theorem \ref{thm: convergegeo} to be of practical use. 
In the next section, we provide a work-around to this issue through the use of a semi-parametric bootstrap. 

\subsubsection{Control limits via semi-parametric bootstrap}

In this section we adopt the control chart in the previous section for the purposes of profile monitoring through the use of bootstrapping.
If $\bm{y}^{(-m+1)}, \dots, \bm{y}^{(0)}$ are the observed responses from $m$ historical profiles, we partition these into sets of size $m - m^\star$ and $m^\star$. 
These two sets of historical profiles are used to create two sets of moment estimates: 
$(\widehat{\bm{\mu}}_{\textrm{monitor}}, \widehat{\bm{\Sigma}}_{\textrm{monitor}})$ using $m-m^\star$ of the historical profiles and $(\widehat{\bm{\mu}}_{\textrm{boot}}, \widehat{\bm{\Sigma}}_{\textrm{boot}})$ using $m^\star$ of the historical profiles. 
We compute $m^\star$ monitoring statistics $\{\widetilde{p}_{\,\cdot\,, i}\}_{i=1}^{m^\star}$ where $\widetilde{p}_{\,\cdot\,, i}$ is one of the two 
monitoring statistic p-values described described in Section \ref{ssec: monitoring statistics}
and computed 
using $(\widehat{\bm{\mu}}_{\textrm{monitor}}, \widehat{\bm{\Sigma}}_{\textrm{monitor}})$. 
Since the historical profiles $\{\bm{y}^{(t)}\}_{t=-m+1}^0$ are independent, the two sets of moment estimates 
$(\widehat{\bm{\mu}}_{\textrm{monitor}}, \widehat{\bm{\Sigma}}_{\textrm{monitor}})$
and 
$(\widehat{\bm{\mu}}_{\textrm{boot}}, \widehat{\bm{\Sigma}}_{\textrm{boot}})$
are independent. 
Therefore, $\{\widetilde{p}_{\,\cdot\,, i}\}_{i=1}^{m^\star}$ are also independent. 

According to 
Theorem \ref{thm: main},
if the lower control limit is set to the $k^{\text{th}}$ order statistic of set of monitoring statistics 
$\{\widetilde{p}_{\,\cdot\,, i}\}_{i=1}^{m^\star}$,
then $\text{ARL}_0 = \frac{m^\star}{k-1}$. 
Although the result on the $\text{ARL}_0$ is exact, the variability of the in-control run lengths (and therefore the FAR) can be impractically large if $m^\star$ is too small. 
\cite{willemain1996CC_EmpiricalRefDist} observed the variability of the in-control run lengths is greater than that of a geometric distribution. 
Conversely, Theorem \ref{thm: convergegeo} showed as $m^\star\to\infty$, the distribution of in-control run lengths converge in distribution to a geometric distribution with parameter $1/\text{ARL}_0$ provided $k = 1 + \frac{m^\star}{\text{ARL}_0}$. 
Due to these considerations, large values of $m^\star$ are ideal. 
However, even if the size of the historical data set $m$ is large, setting $m^\star$ to a large value runs the risk of there being an insufficient number of historical profiles to estimate the first two moments of the in-control process. 
We address this challenge by using bootstrapping to artificially increase the value of $m^\star$ without decreasing the ability to estimate the in-control process. 

We aim to compute our control limit on a set of $m^\prime > m$
bootstrapped historical profiles. 
Under this setup, 
the control limit will be the $\left(\frac{m^\prime}{\text{ARL}_0} + 1\right){\text{th}}$ 
order statistic of the $m^\prime$ bootstrapped profiles,
analogous to the non-bootstrapped case described in the previous paragraph. 
To avoid 
interpolation between order statistics, we must ensure $\frac{m^\prime}{\text{ARL}_0} + 1$ is an integer. 
Therefore, we choose values of $b_1, b_2 \in \mathbb{Z}^+$ in the following procedure and set $m^\prime = b_1 \, b_2 \, \text{ARL}_0$. 

To account for variability in the moment estimates $(\widehat{\bm{\mu}}_{\textrm{boot}}, \widehat{\bm{\Sigma}}_{\textrm{boot}})$, 
we perform a parametric bootstrap 
to generate $b_1$ sets of $b_2 \, \text{ARL}_0$ bootstrapped in-control profiles as follows: 
\begin{itemize}
        \item For each $j = 1,\dots, b_1$,
         do the following: 
        \begin{itemize}
            \item Generate $\bm{y}^{\star,(-m+1)}_j, \dots, \bm{y}^{\star,(0)}_j \sim N(\widehat{\bm{\mu}}_{\textrm{boot}}, \widehat{\bm{\Sigma}}_{\textrm{boot}})$
            \item Compute $\widehat{\bm{\mu}}_j$ and $\widehat{\bm{\Sigma}}_j$ from the above generated  dataset. 
            \item Generate $B$ in-control profiles from the $N(\widehat{\bm{\mu}}_j,  \widehat{\bm{\Sigma}}_j)$ distribution. 
        \end{itemize}
    \end{itemize}
For the $b_1 \, b_2$ sets of bootstrapped in-control profiles, we compute $b_1 \, b_2$ monitoring statistics where the conditional p-values are computed using         $\widehat{\bm{\mu}}_{\textrm{monitor}}$ and $\widehat{\bm{\Sigma}}_{\textrm{monitor}}$.
The next section describes how our choice of monitoring statistic and control limit compares to competitors. 

\section{Simulation studies}\label{sec:simulations}

We conduct simulation studies to confirm the results in Theorems \ref{thm: convergegeo} and to assess the strengths and weaknesses of our proposed methodology. 
As mentioned in Section \ref{sec:background}, the closest competitors in methodology are \cite{Colosimo2010Comparison} and \cite{li2023TweedieProfmon}. 
We compare our proposed approach against that of \cite{Colosimo2010Comparison} and herein after call this the PCA control chart. 
We omit comparisons with \cite{li2023TweedieProfmon} in the simulation section as all of the simulation setups violate their assumption of independent within-profile increments. 
As the location of the monitors are fixed, this approach can be compared to more classical multivariate control charts such as the Phase II Hotelling $T^2$ chart using estimates for the mean vector and covariance matrix using $m$ historical observations.

\subsection{Verifying Theorem \ref{thm: convergegeo} via simulation}
\begin{figure}[ht]
    \centering
    \subfloat[Standard Deviation of Run Lengths]{
    \includegraphics[width = 0.32\linewidth]{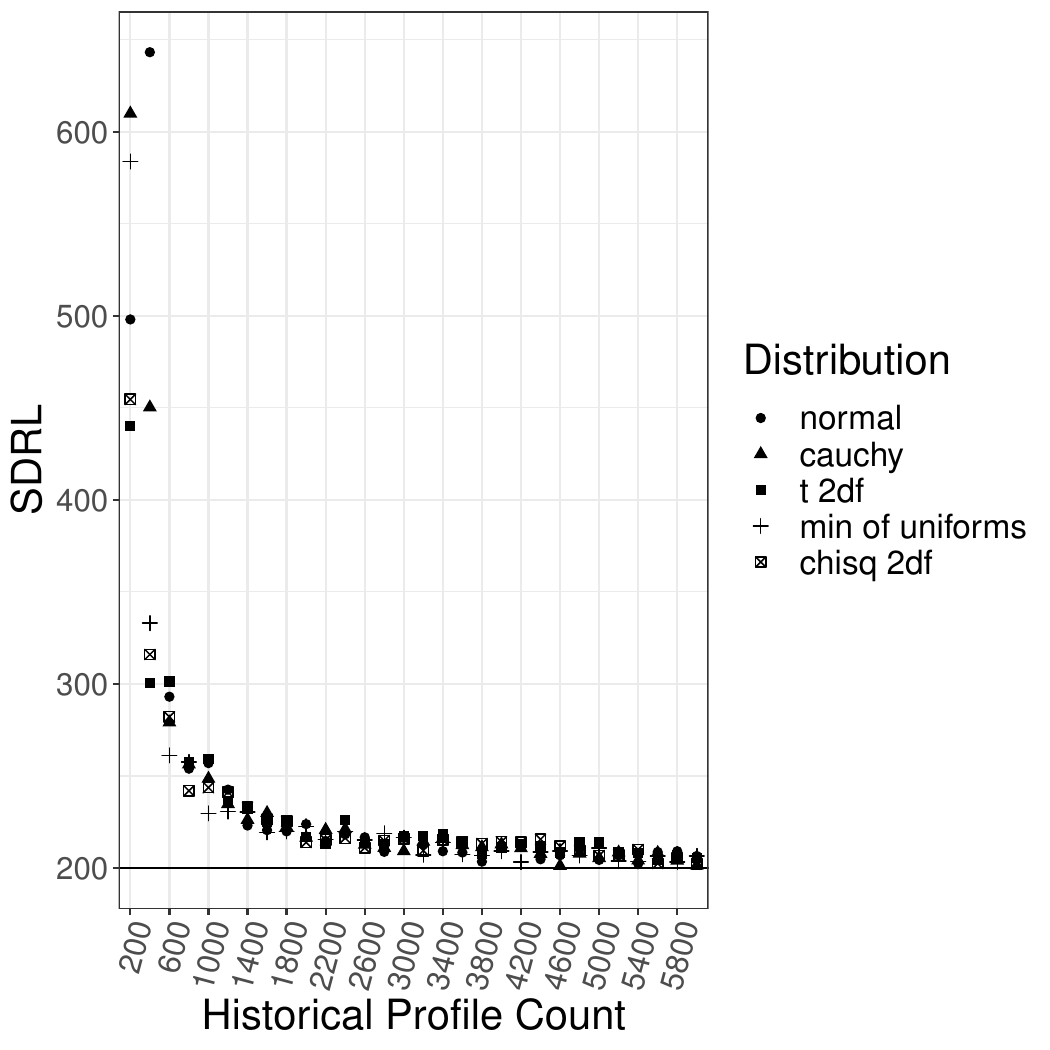}
    \label{fig:SDRL}
    }
    \subfloat[Sample ARLs (Normal)]{
    \includegraphics[width = 0.32\linewidth]{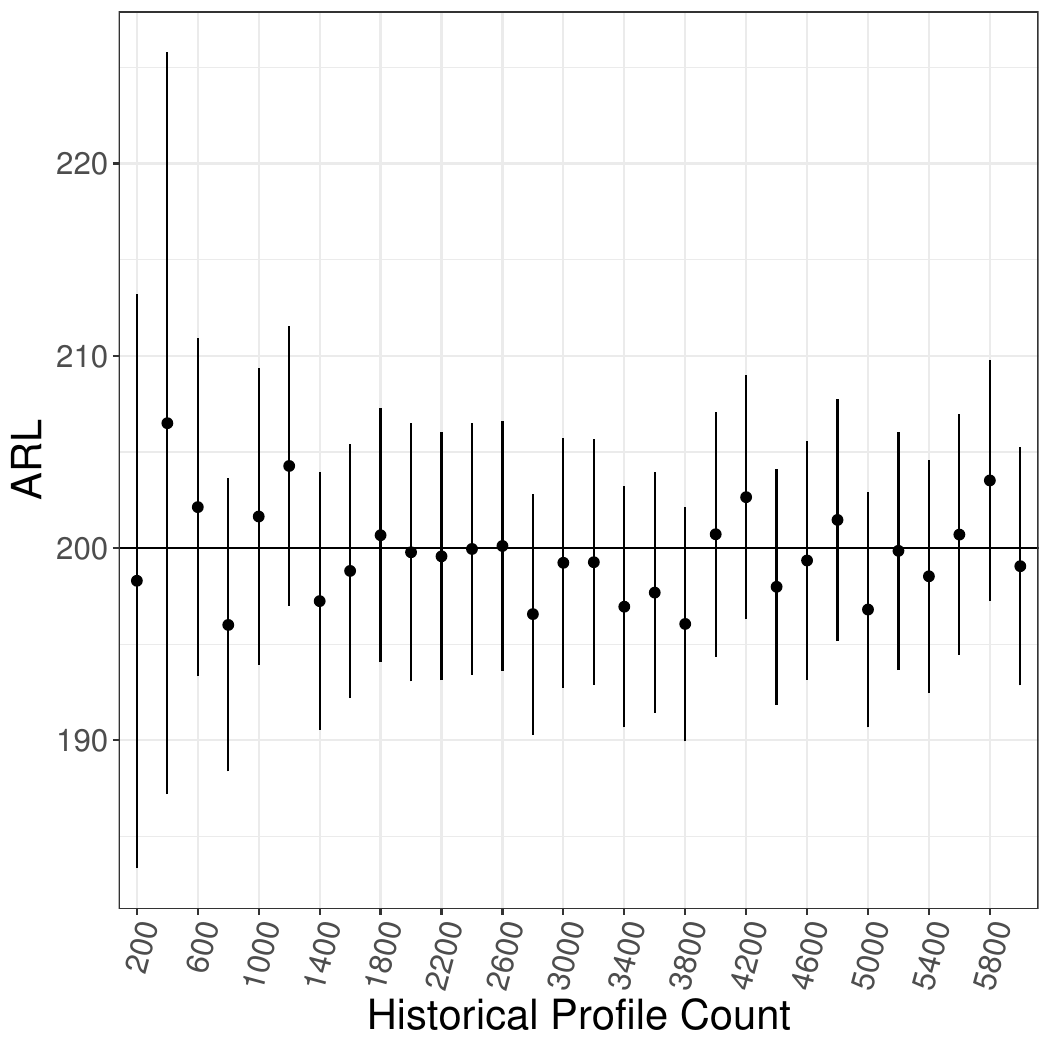}
    \label{fig:ARLnormal}
    }
    \subfloat[Sample ARLs ($t_2$)]{
    \includegraphics[width = 0.32\linewidth]{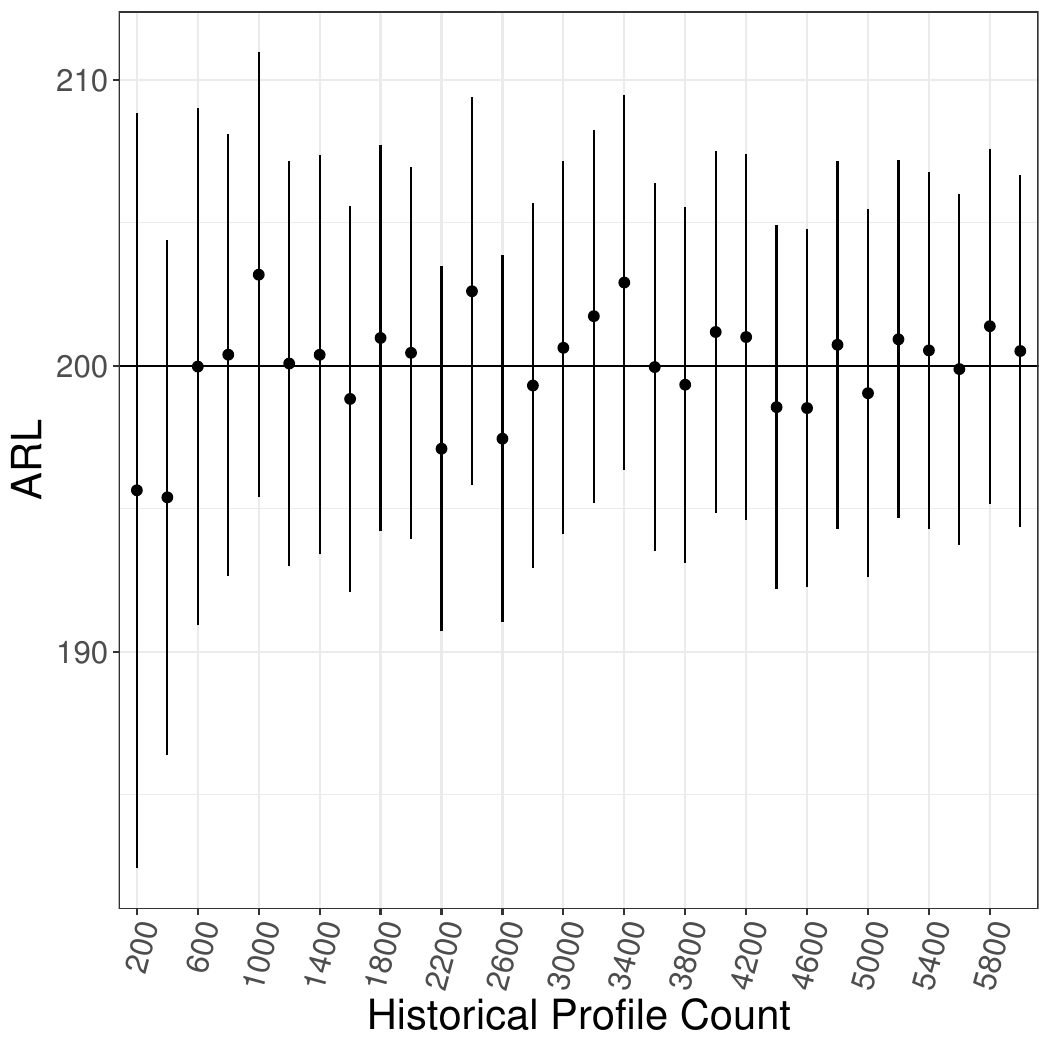}
    \label{fig:ARLt}
    }
    \\
    \subfloat[Sample ARLs (Cauchy)]{
    \includegraphics[width = 0.32\linewidth]{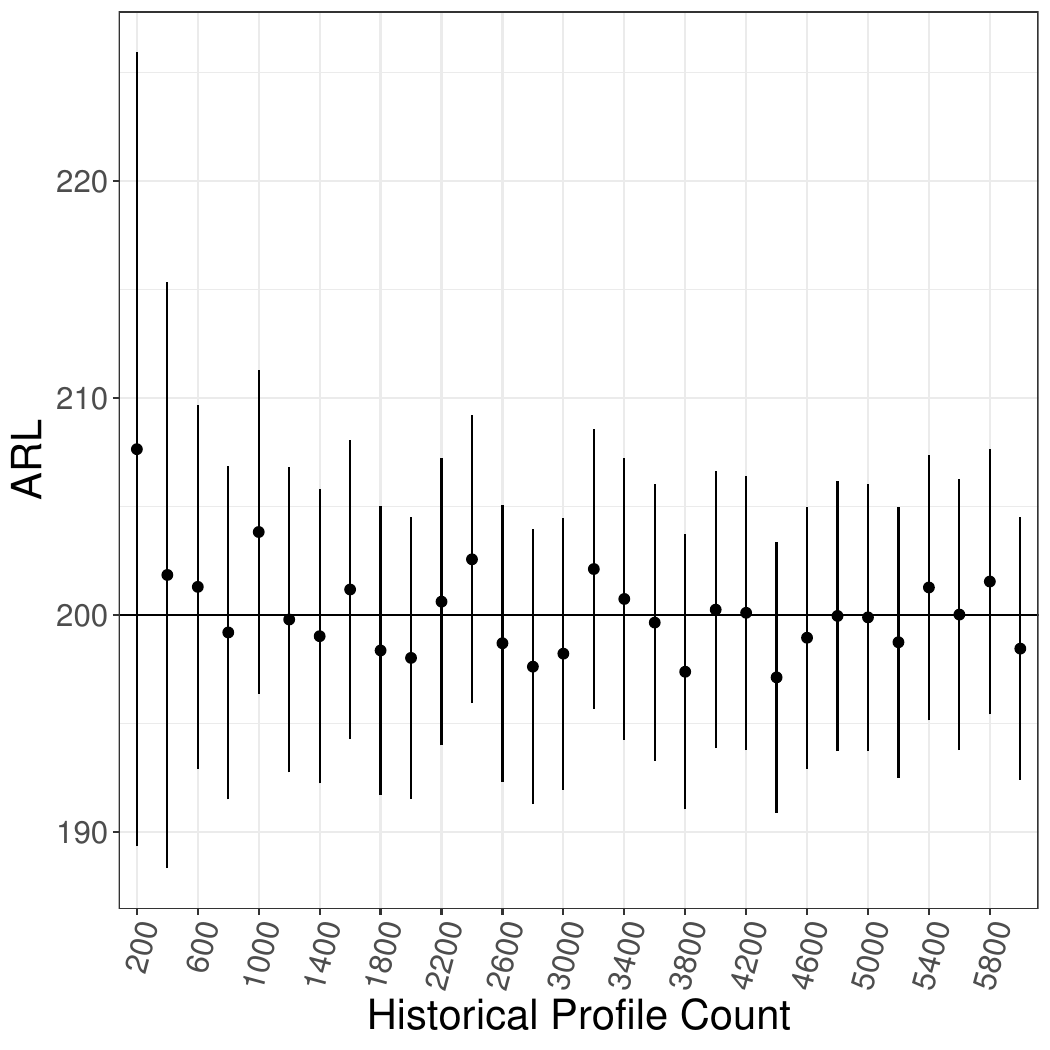}
    \label{fig:ARLcauchy}
    }
    \subfloat[Sample ARLs ($\chi^2_2$)]{
    \includegraphics[width = 0.32\linewidth]{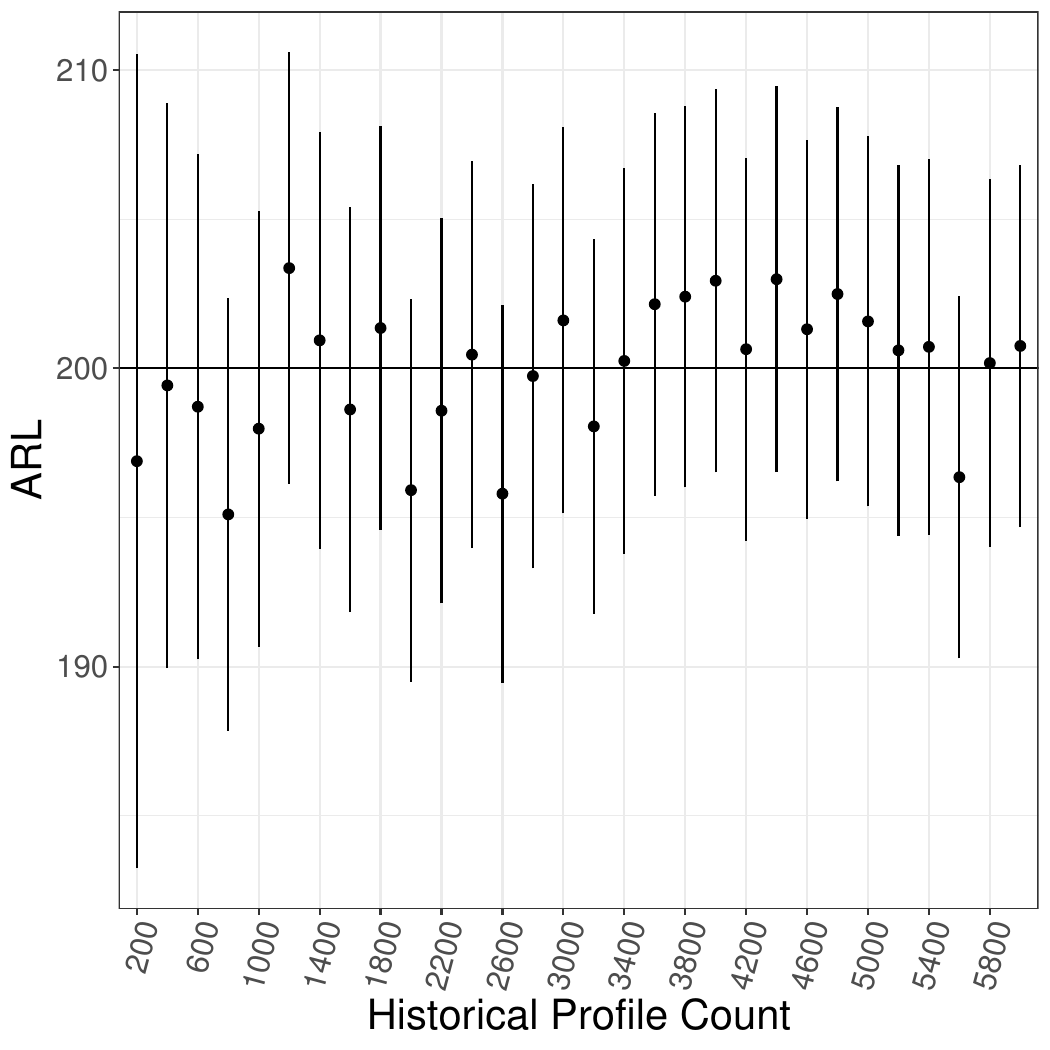}
    \label{fig:ARLchisq}
    }
    \subfloat[Sample ARLs (Beta(1, 10))]{
    \includegraphics[width = 0.32\linewidth]{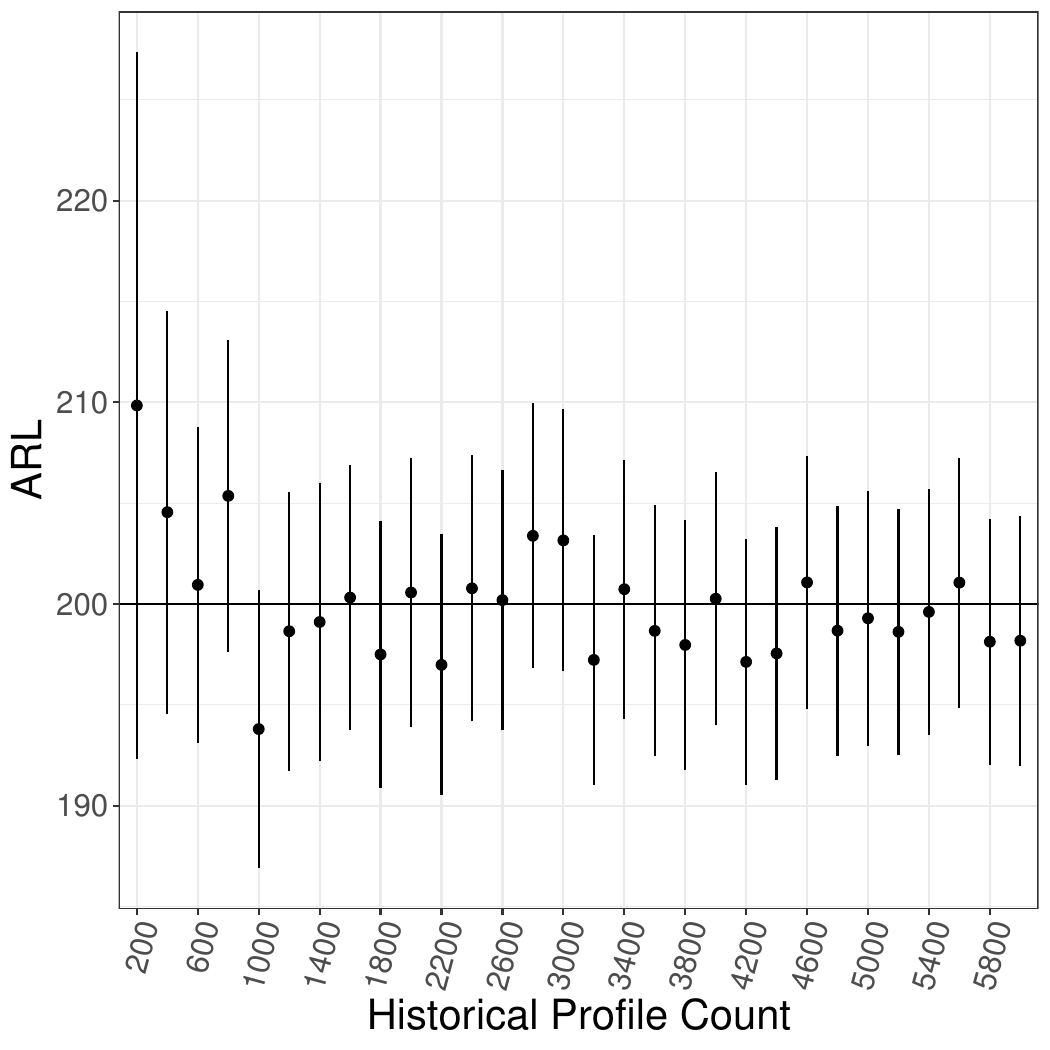}
    \label{fig:ARLminunif}
    }
    \caption{Results of the simulation study verifying Theorems \ref{thm: main} and \ref{thm: convergegeo}. 
    }
    \label{fig: simulations}
\end{figure}

We simulate the in-control run lengths of the control chart in Theorem \ref{thm: main} to empirically confirm the expression for the $\text{ARL}_0$ is correct and the first two moments of the run lengths converge to that of a Geometric random variable.
To show our control chart does not depend on the underlying distribution of the data process, we chose five different distributions from which to simulate: 
\begin{enumerate}[itemsep=0em,topsep=.1em]
    \item {\bf Normal distribution:} $N(0,1)$.
    \item {\bf $t$-distribution:} $t_{\nu}$ with $\nu = 2$ degrees of freedom.
    \item {\bf Cauchy:} zero location parameter and unit scale parameter
    \item {\bf Chi-Square:} $\chi_{k}^2$ with $k = 2$ degrees of freedom. 
    \item {\bf Beta:} $\text{Beta}(1, \, 10)$. 
\end{enumerate}
The $t$-distribution was chosen as it has a finite mean, 
but an infinite variance, and the Cauchy was chosen because none of the moments of the Cauchy distribution exist. 
The Beta distribution was chosen as it models the minimum of 10 uniformly distributed random variables, and lastly the Chi-square distribution was chosen as it is the distribution of the monitoring statistic from a profile monitoring approach on simple linear models \citep{Kang2000}. 
For each distribution, we conducted $10,000$ replications by simulating the historical dataset $U_1, \ldots, U_m$ and the online monitoring data $V_1, V_2, \ldots$ under the in-control process. 
The value of $k$ to set the lower control limit was chosen to have $\text{ARL}_0 = 200$.
The values of $m$ were chosen to be multiples of 200 to ensure $k$ is an integer, avoiding the need to round or approximate to the nearest integer value. 
We present our results in Figure \ref{fig: simulations}.

\subsection{Simulation studies studying the profile monitoring method}
We conduct three simulation studies, each consisting of a set of scenarios. 
Each scenario was replicated $1,000$ times.
In each replication, a set of $m=1,000$ historical and in-control profiles were randomly generated. 
For the proposed control charts, the appropriate control limit was computed from these $m$ profiles and the desired $\text{ARL}_0$ of $1,000$. 
Control limits for competing control charts were calibrated to achieve the same FAR as was observed for our control charts. 
Once control limits were established, monitoring began wherein a new profile was randomly generated for each time step. 
In generating both historical and online profiles, the location of the monitor sites (i.e., predictors of a profile) were held fixed for all time-steps. 
The times of the false alarms were recorded for each replication. 
Monitoring ended when either a true alarm was raised or if $25,000$ time steps passed since monitoring began. 
If $t^{\star}_{i}$ denotes the time when a true alarm occurs for the $i^{\text{th}}$ trial, the Monte Carlo estimate of 
$\text{ARL}_1$ is 
$\frac{1}{1000} \sum_{i=1}^{1000} (t^{\star}_{i} - \tau)$, assuming that all 1000 trials raised a true alarm before time step $25,000$ and denoting the true time at which the 
process goes out-of-control by $\tau$. 
Assuming all replications of a scenario end in a true alarm, the estimate of the FAR is the proportion of alarms that are false alarms. 

The in-control and out-of-control processes are summarized in Table \ref{tbl:ICOOCprocesses}.
In all cases $n=10$ monitor sites are used and are chosen to be equispaced.   
The set of additive errors $\{\epsilon_x\}_{\{x\in \mbX\}}$ are independent and identically distributed  $N(0, 0.1^2)$.
In all simulation studies, a parameter $\xi\in \{0, 0.1, \dots, 1\}$ is chosen to control how different the out-of-control profile is from the in-control profile. 
A value of $\xi=0$ makes the out-of-control profile identical to the in-control profile and a value of $\xi = 1$ denotes a notion of `maximal' difference.
For determining control limits as described in the previous section, $b_1= 100$, $m^\star = m/2$, and $b_2$ is varied between 5, 10, and 20. 

\begin{table}[t]
\centering
\scalebox{0.70}{
\begin{tabular}{llllll}
\hline
\begin{tabular}[c]{@{}l@{}}Simulation \\ Study\end{tabular} & \begin{tabular}[c]{@{}l@{}}In-control \\ Process\end{tabular}                                                                            & \begin{tabular}[c]{@{}l@{}}Out-of-Control\\ Process\end{tabular}                                                                                                                                      & $\mbX$ & $\alpha$ & $Z$    \\ \hline
1                                                           & $\alpha \sin(x) + \epsilon_x$                                                                                                            & $\alpha \sin(x) + \xi + \epsilon_x$                                                                                                                                                                   & $[0.1, 2\pi-0.1]$  & $N(0,1)$   & -      \vspace{.8cm} \\
2                                                           & $\alpha \sin(x) + \epsilon_x$                                                                                                            & \begin{tabular}[c]{@{}l@{}}$\begin{cases}\alpha \, \sin(x) + \epsilon_x & \mbox{if } x \neq x_3 \\ (1-\xi)\left[ \alpha \, \sin(x) + \epsilon_x \right] +\xi \, Z & \mbox{if } x = x_3 \end{cases}$\end{tabular} & $[0.1, 2\pi-0.1]$  & $N(0,1)$   & $N(0,1)$ \vspace{.8cm} \\
3                                                           & 
\begin{tabular}[c]{@{}l@{}} $f(x, \bm{\alpha}) + \epsilon_x$ \\ 
where $f(x, \bm{\alpha}) = \dsum_{k=0}^{6} \, \alpha_k \, x^k$
\end{tabular} 
& \begin{tabular}[c]{@{}l@{}}$\begin{cases} f(x, \balpha) + \epsilon_x & \mbox{if } x \leq x_6 \\ - (f(x,\balpha) - f(x_6, \balpha)) + f(x_6, \balpha) + \epsilon_x & \mbox{if } x > x_6.  \end{cases}$\end{tabular}    
& $[0, 1]$     & $N(\xi,1)$   & -     
\end{tabular}
}
\caption{In-control and Out-of-control processes used in simulation studies 1-3}
\label{tbl:ICOOCprocesses}
\end{table}

\subsection{Simulation study 1: Global shift}

\begin{figure}[t]
\centering 
\includegraphics[width = \linewidth]{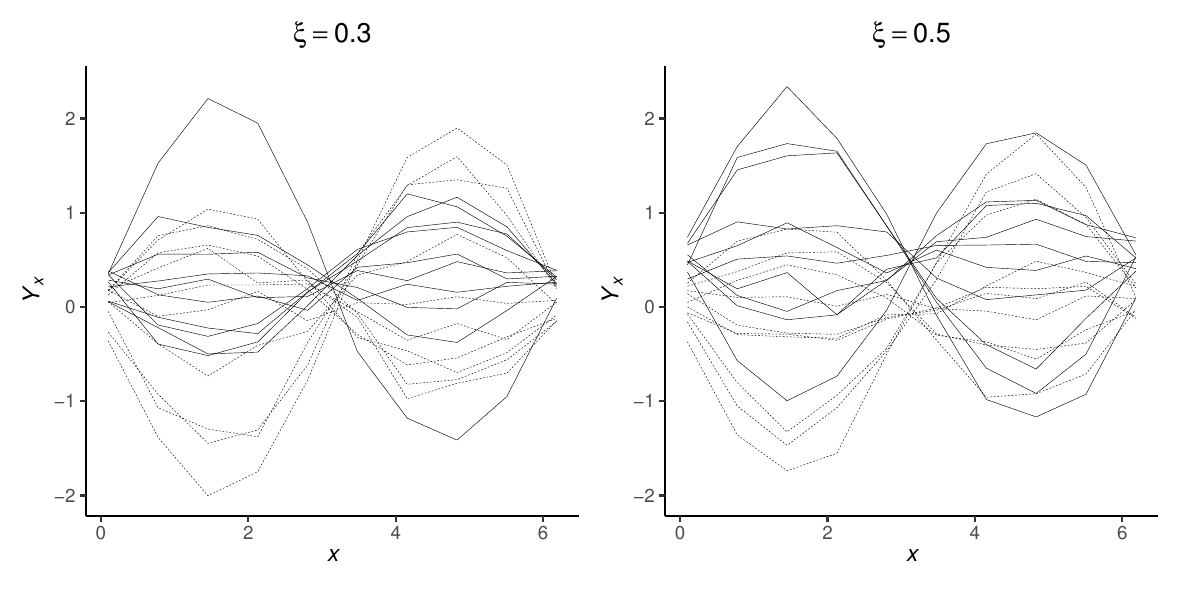}
\caption{\label{fig:sim1_ex} Visualizations of the in-control (dashed) and out-of-control (solid) 
processes in Simulation Study 1 for two different shifts $\xi \in \{.3, .5\}$.}
\end{figure}

We revisit the motivating example of the function $f(x, \alpha) = \alpha \, \sin(x)$ ($x \in [0, \, 2 \pi]$) 
from Section \ref{sec:intro}.
See Table \ref{tbl:ICOOCprocesses} for specification of the in-control and out-of-control processes.
The motivation to not include the end points $0$ and $2 \pi$ of the interval $[0, \, 2 \pi]$ 
is due to the fact that 
$\sin(0) = 0 = \sin(2 \, \pi)$ which would result in $Y_0 = \epsilon_0$ and $Y_{2 \pi} = \epsilon_{2 \pi}$. 
Thus, 
we choose only interior points for which $\alpha \, \sin(x)$ is non-degenerate.

Consider an out-of-control process denoted by $\{W_x\}_{x \in \mbX}$ where 
\beno
W_x 
\= \alpha \, \sin(x) + \epsilon_x + \xi,
&& x \in \mbX, 
&\text{for } \xi \in \{0, .1, .2, \ldots, .9, 1.0\}, 
\ee
where the value $\xi$ shifts or translates the entire functional response. 
We visualize the in-control and out-of-control processes at two different shifts $\xi \in \{.3, .5\}$ 
in Figure \ref{fig:sim1_ex}. 
\begin{table}[t]
\centering
\begin{tabular}{r|rrrrrrrrrr}
  \hline
 Shift & $x_1$ & $x_2$ & $x_3$ & $x_4$ & $x_5$ & $x_6$ & $x_7$ & $x_8$ & $x_9$ & $x_{10}$ \\
  \hline
  .1 & 0.71 & 0.14 & 0.10 & 0.12 & 0.29 & 0.29 & 0.12 & 0.10 & 0.14 & 0.71 \\
  .2 & 1.42 & 0.28 & 0.20 & 0.23 & 0.58 & 0.58 & 0.23 & 0.20 & 0.28 & 1.42 \\
  .3 & 2.12 & 0.42 & 0.30 & 0.35 & 0.87 & 0.87 & 0.35 & 0.30 & 0.42 & 2.12 \\
  .4 & 2.83 & 0.57 & 0.40 & 0.47 & 1.16 & 1.16 & 0.47 & 0.40 & 0.57 & 2.83 \\
  .5 & 3.54 & 0.71 & 0.50 & 0.58 & 1.44 & 1.44 & 0.58 & 0.50 & 0.71 & 3.54 \\
  \hline 
\end{tabular}
\caption{\label{tab:sim1_snr} Signal-to-noise ratios at each of the $10$ monitoring sites across shifts.} 
\end{table}
We quantify the difficulty of the detection problem through computing the signal-to-noise ratio at each of the monitor sites. 
Note that based on the specification of the in-control and out-of-control processes, 
the local-site SNR can be quantified as 
\beno
\text{SNR}(x) 
\= \dfrac{\mbE(Y_x - W_x)}{\text{SD}(Y_x)}
\= \dfrac{\xi}{\sqrt{\sin(x)^2 + .1^2}}.   
\ee
We compute this at each of the ten monitor times $\bx = (x_1, \ldots, x_{10})$ 
and for each shift $\xi \in \{.1, .2, \ldots, .5\}$ in Table \ref{tab:sim1_snr}.  
Through this, we see that the largest local-site SNR occurs at time points $x_{1}$ and $x_{10}$. 
This is due to the fact that the standard deviation will be smallest at those two time steps 
(owing to the fact that $\sin^2(x)$ will be closest to $0$ at those two monitor times).  
We can interpret the local-site SNR as providing information on how challenging the detection problem in this simulation is based on a mean shift of two normal populations,
where the SNR is quantifying the number of standard deviations the shifted mean of the out-of-control process is away from the true mean under the in-control process. 
Table \ref{tab:sim1_snr} provides additional insight to some of the unique aspects of profile monitoring of random functions. 
Not all monitor sites $x \in \mbX$ will provide the same information. 
Indeed, looking across columns, the variation of local-site SNR values varies across different monitoring sites.  

\begin{figure}[h!]
\centering
% \subfloat[blah]{
\subfloat[][\cite{Colosimo2010Comparison}]{
\includegraphics[width=0.47\linewidth]{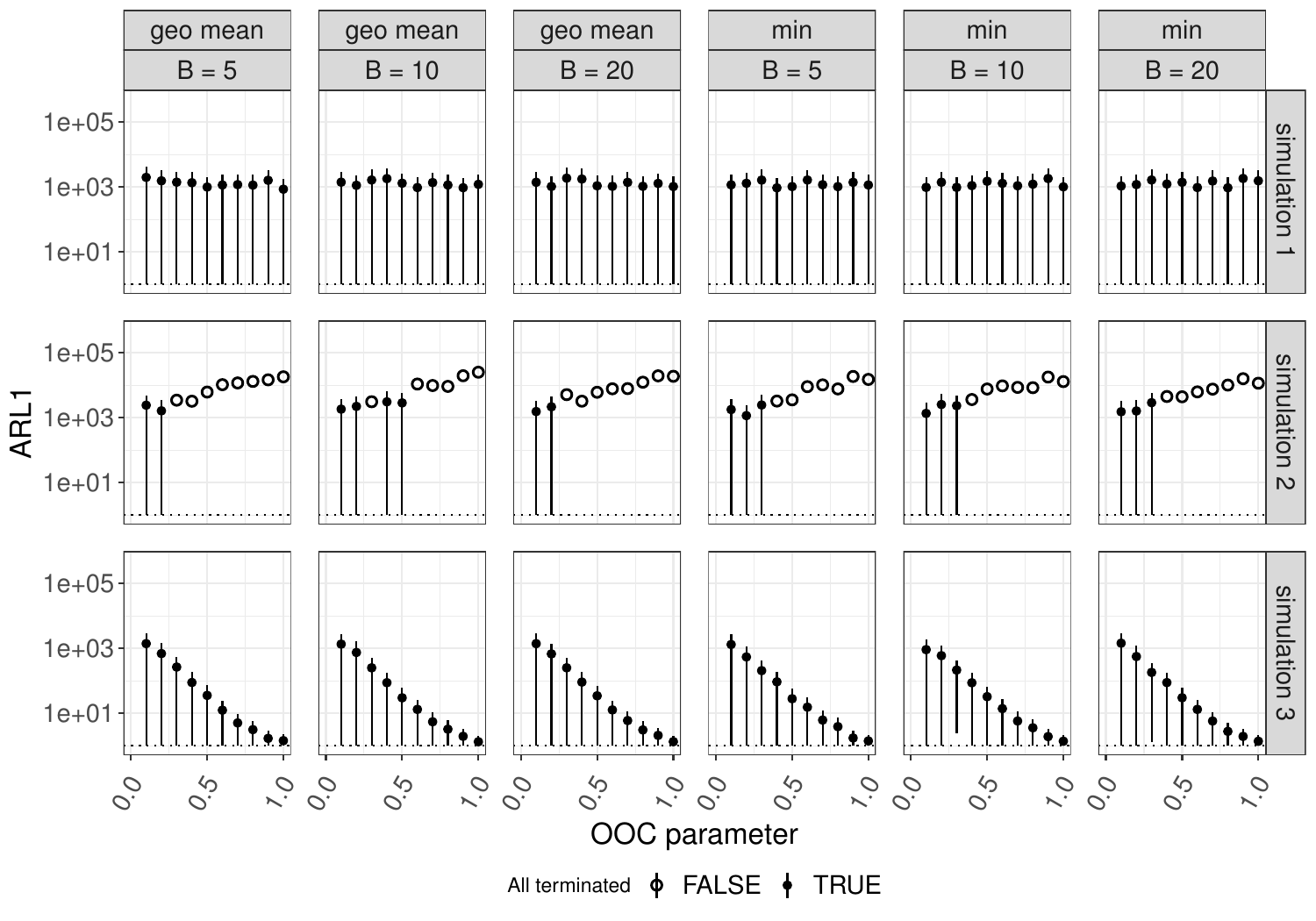}
\label{fig:arl1_pca}
}
\,\,
% \\
\subfloat[][Hotelling $T^2$]{
\includegraphics[width=0.47\linewidth]{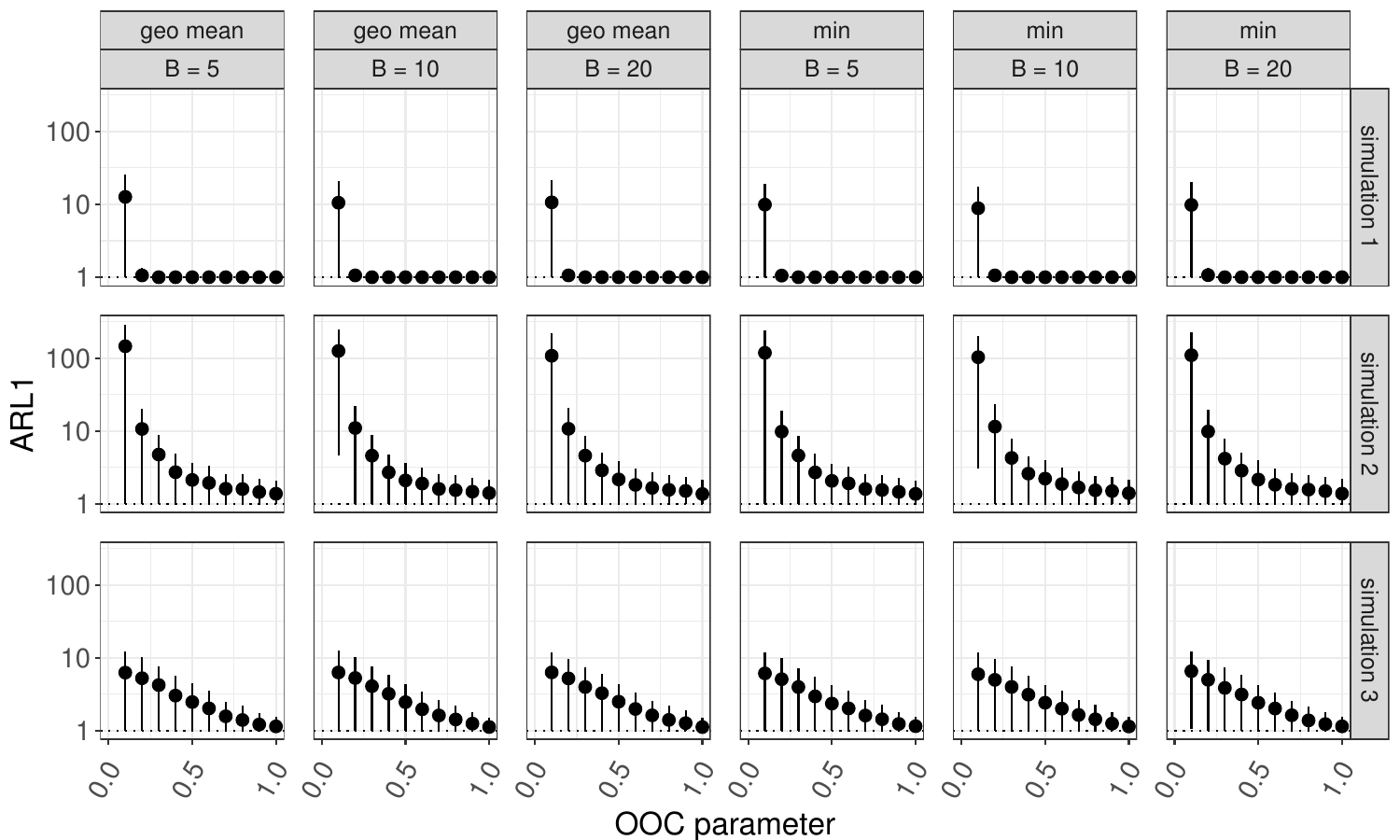}
\label{fig:arl1_t2}
}
\\
\subfloat[][Proposed Control Chart vs. Hotelling $T^2$]{
\includegraphics[width=0.9\linewidth]{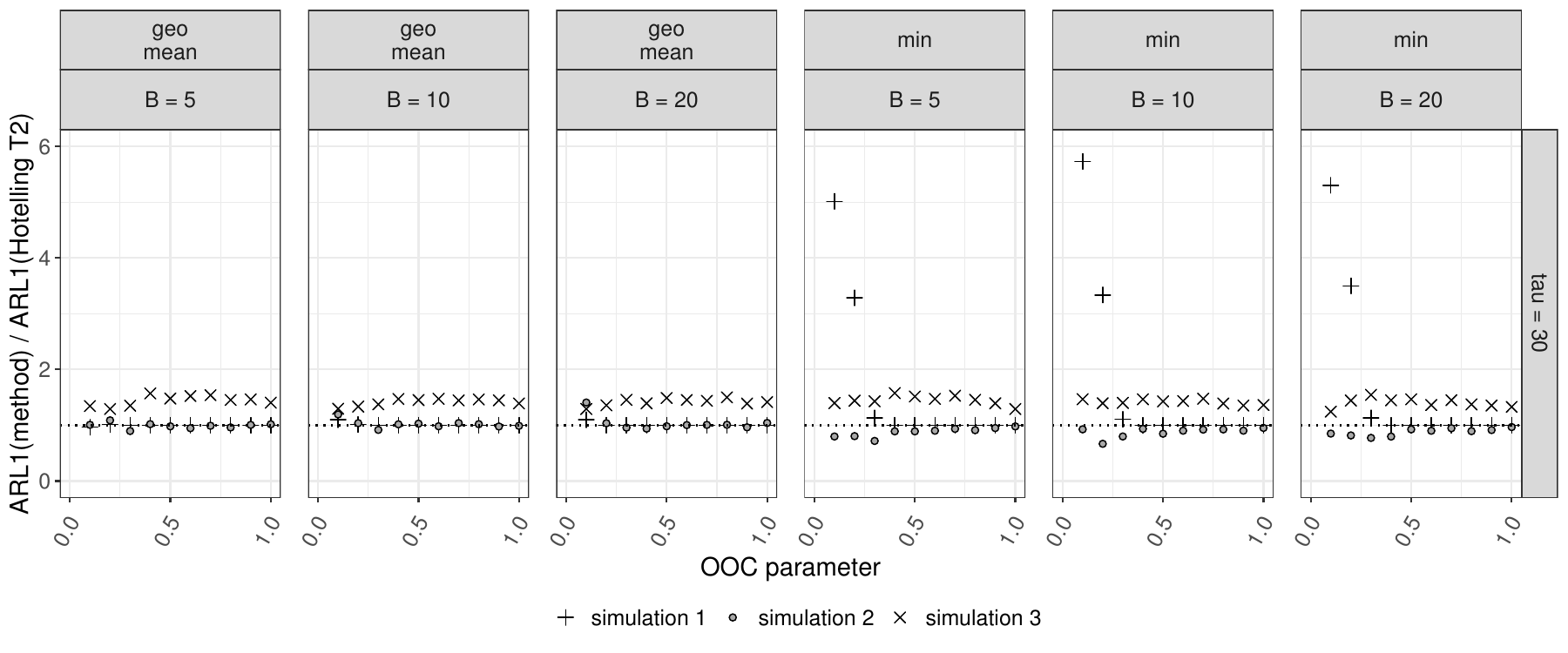}
\label{fig:arl1_simall}
}
\caption{
Sample $\text{ARL}_1$ from simulation studies 1-3.
(a-b) The point corresponds to the estimate of $\text{ARL}_1$ and the length of the vertical line above and below the point is one standard deviation of the out-of-control run lengths.
Due to the log scale, we truncate at the smallest possible run length. 
In (a), not all trials terminated with an alarm after 25000 timesteps, so a lower bound on $\text{ARL}_1$ is provided.
(c) The sample $\text{ARL}_1$ from the proposed control chart is compared against the performance of the Hotelling $T_2$  
}
\label{fig:arl1}
\end{figure}

\begin{figure}[h!]
\centering
\includegraphics[width=\textwidth]{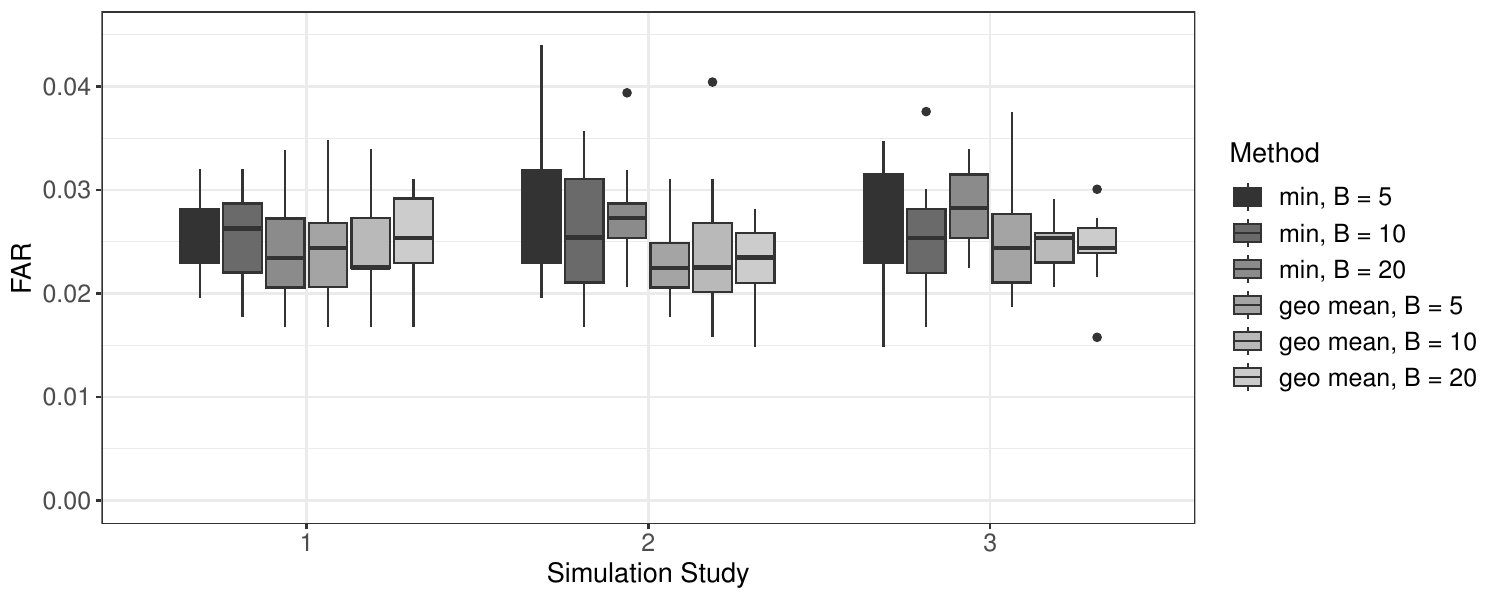}
\caption{FARs from Simulation Studies 1-3.
}
\label{fig:FAR}
\end{figure}

Figure \ref{fig:arl1_simall} compares the $\text{ARL}_1$ of the Hotelling $T^2$ control chart against both of our proposed control charts.
Our control chart performs just as well as the Hotelling $T^2$ control chart, 
aside from very small effect sizes of $\xi \in \{0.1, 0.2\}$ when using the minimum rule. 
As the method of aggregating conditional p-values by taking the minimum ignores information from all but one of the monitors, it performs worse than the method of aggregating conditional p-values using the geometric mean. 
This makes sense as the global mean shift is captured by all of the monitors. 
Interestingly, Figure \ref{fig:arl1_pca} shows the PCA control chart is unable to detect a mean shift. 
The FARs as shown in Figure \ref{fig:FAR} demonstrate all versions of our control chart have comparable FARs in this simulation study. 
\subsection{Simulation study 2: Broken monitor}
\begin{figure}[t]
\centering 
\includegraphics[width = \linewidth]{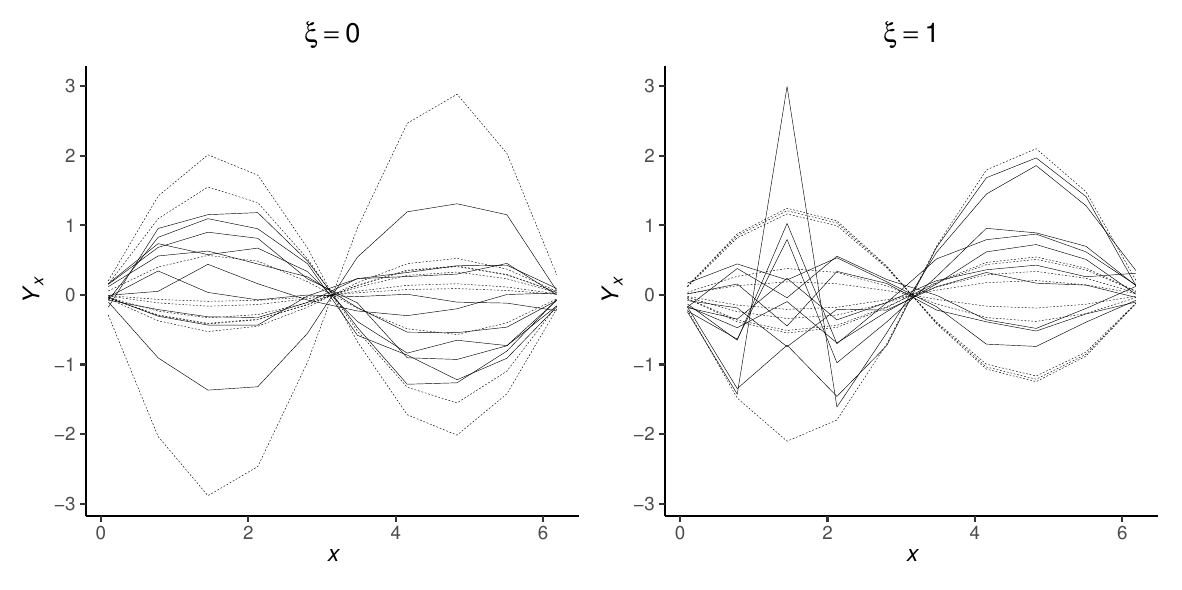}
\caption{\label{fig:sim2_ex} Visualizations of the in-control (dashed) and out-of-control (solid) 
processes in Simulation Study 2 for $\xi \in \{0, 1\}$.}
\end{figure}

We continue with the motivating example of the function 
$f(x, \alpha) = \alpha \, \sin(x)$ ($x \in [0, \, 2 \pi]$) 
from Section \ref{sec:intro}.  
As described in Table \ref{tbl:ICOOCprocesses} in-control process remains unchanged from the previous simulation study and the out-of-control processes is modified.
We visualize the in-control and out-of-control processes at two values $\xi \in \{0, .5\}$
in Figure \ref{fig:sim2_ex}.
The out-of-control process may be interpreted as a broken monitor,  
i.e,
one in which the true signal is contaminated with differing levels of noise.
The goal in this simulation study is therefore to detect when the signal is contaminated by noise and thus 
not reflecting the true signal,
which in this case is the functional behavior.
Moreover, the we highlight that there is no mean shift for any of the monitors, but the change is purely in the covariance structure of ten monitors. 
Figure \ref{fig:sim2_ex} demonstrates how the functional behavior at monitor $x_3$ begins to degrade 
as $\gamma$ becomes smaller (which corresponds to the observed process being more noise than signal).    

As seen in Figure 6, the sample FARs become less variable as the number of bootstrap samples increases (i.e., as $b_2$ increases), especially when using the minimum rule. 
Figure \ref{fig:arl1_simall} compares the $\text{ARL}_1$ of the Hotelling $T^2$ control chart against both of our proposed control charts.
Contrary to the global mean shift example, our control chart using the minimum rule is quicker to raise a true alarm. 
Figure \ref{fig:arl1_pca} shows the PCA control chart is unable to detect the out-of-control process. 
In fact, as the effect size grows, the PCA control chart actually takes longer to raise a true alarm. 
As not all of the trials terminated with an alarm by timestep 25,000, a lower bound of $\text{ARL}_1$ is provided for the PCA control chart in Figure \ref{fig:arl1_pca}.

\subsection{Simulation study 3: Trajectory switch}
\begin{figure}[t]
\centering 
\includegraphics[width = \linewidth]{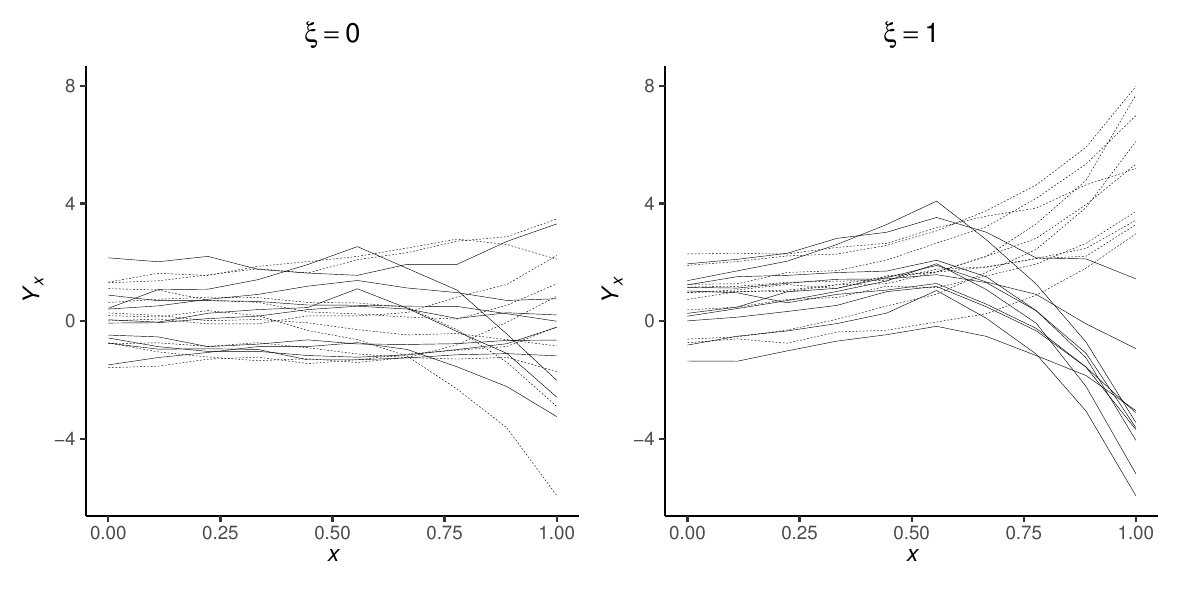}
\caption{\label{fig:sim3_ex} Visualizations of the in-control (dashed) and out-of-control (solid) 
processes in Simulation Study 3.
For $k = 1,\dots, 6$, the basis coefficients $\alpha_k$ are distributed $N(0,1)$ on the left and $N(1,1)$ on the right.
}
\end{figure}
We next study a different basis structure for the in-control process, 
namely that of a monomial basis.
See Table \ref{tbl:ICOOCprocesses} for specifications of the in-control and out-of-control processes.
This example differs from the previous in-control processes by including multiple basis functions.
We visualize the in-control and out-of-control process in Figure \ref{fig:sim3_ex}.
The out-of-control process essentially continues the path of the would-be in-control process, with the deviation of changing direction. 
In other words, if the process would increase after time $x_6$,
it now proceeds to decrease at the same rate, reversing the intended direction. 

As seen in Figure 6, the sample FARs become less variable as the number of bootstrap samples increases (i.e., as $b_2$ increases). 
Figure \ref{fig:arl1_simall} compares the sample $\text{ARL}_1$ for the Hotelling $T^2$ control chart and both of our proposed control charts and shows they are similar with Hotelling $T^2$ having slightly quicker detection.  
For the cost of a slightly longer detection delay, we gain the ability to isolate out-of-control monitors by inspecting the conditional p-values, which is not an ability that is a clear byproduct of using the Hotelling $T^2$ control chart. 
We include a visualization of the conditional p-values for the in-control and out-of-control monitors in the supplementary materials. 
Figure \ref{fig:arl1_pca} shows the PCA control chart is eventually able to detect an out-of-control profile, but it is considerably slower than the other methods considered. 

\section{Real-World Example}
\begin{figure}[th!]
\centering
\includegraphics[width =  \linewidth, keepaspectratio]{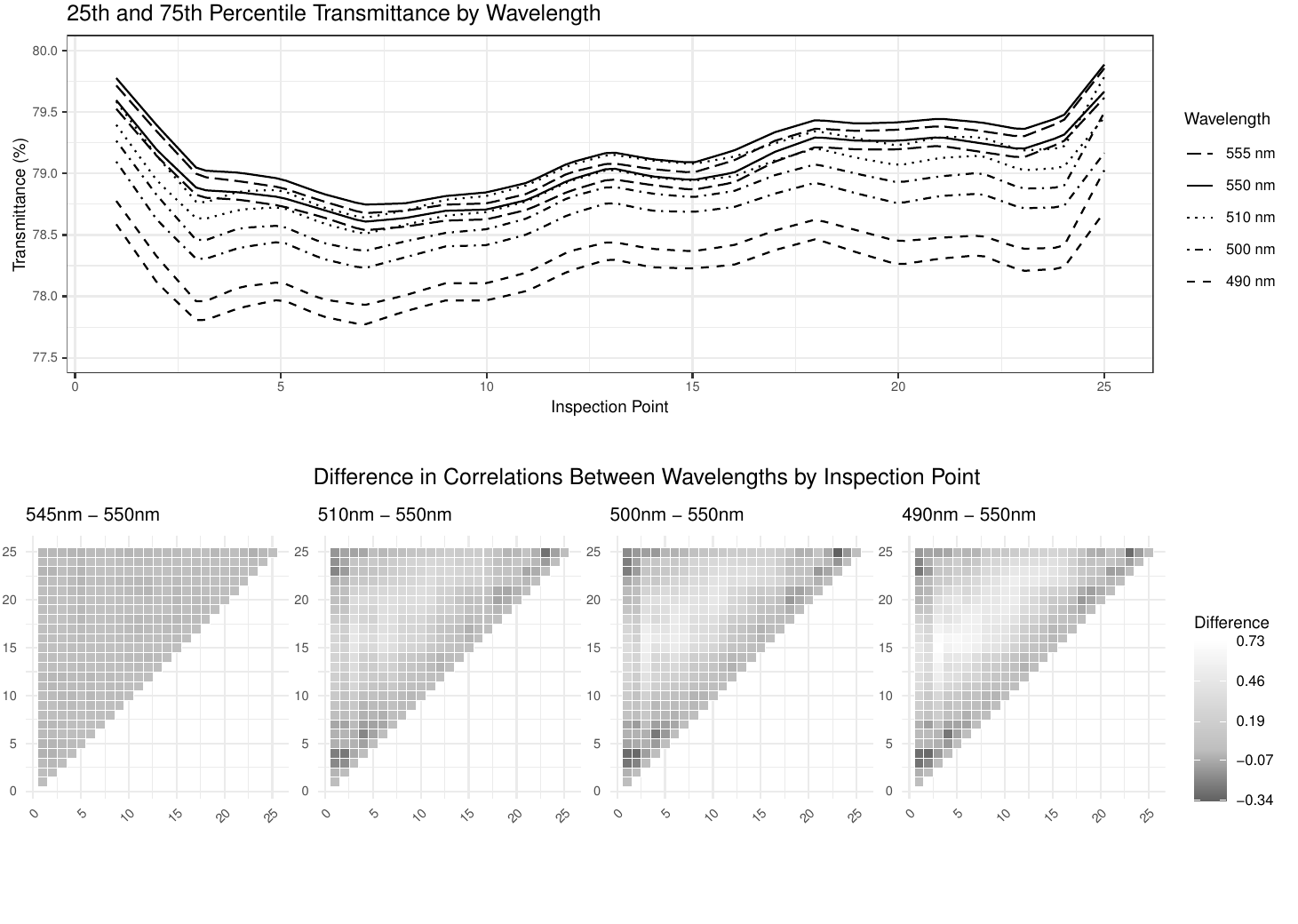}
\caption{\label{fig:lowE} A visualization of transmittance profile across multiple inspection points. 
For each inspection point the 25th and 75th percentile transmittance is plotted and an interpolating piecewise linear fit is displayed for each set of points.
}
\end{figure}
\begin{figure}
    \centering
    \includegraphics[width = \linewidth]{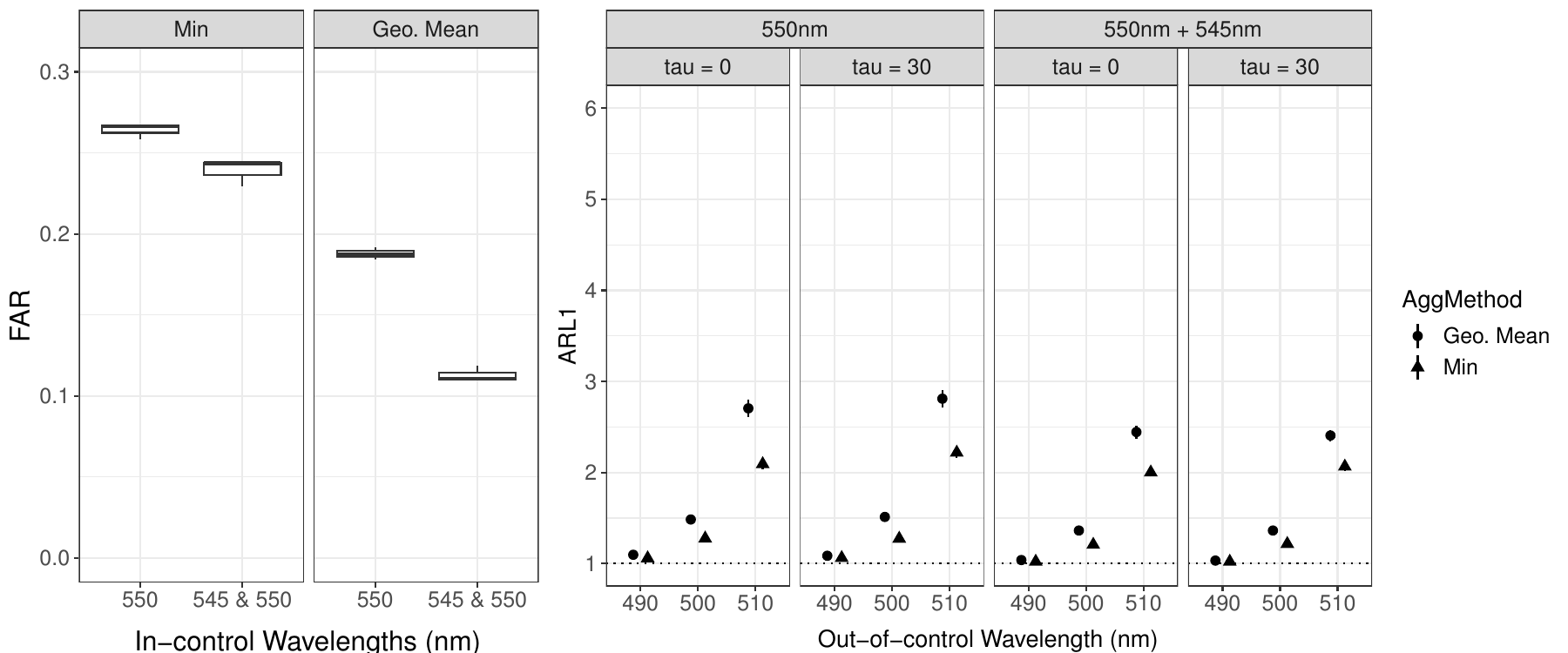}
    \caption{FAR and $ARL_1$ for the low-emittance glass dataset.
    The standard error of the out-of-control run lengths are depicted with vertical lines above and below the estimated $\text{ARL}_1$.}
    \label{fig:ARL1_FAR_lowE}
\end{figure}

In this section we apply the proposed control chart to a dataset on low-emittance (low-E) glass from \cite{ Li2015LowE,Li2018MultivarProfGP}.
Low-E glass is designed to ``lower heat flow through a window by reflecting 90\% of infrared radiation, while still allowing visible light to enter'' \cite{Li2015LowE}. 
In this dataset a collection of 192 glass panels are tested at 25 inspection locations with 35 different wavelengths. 
At each inspection site, the transmittance for a particular wavelength is measured. 
Figure \ref{fig:lowE} shows transmittance profiles at different wavelengths for all 192 glass panels. 
We take one one transmittance profile at wavelength of 550nm to be the in-control profile. 
We choose three other wavelengths' transmittance profiles to act as out-of-control profiles to imitate different different amounts of signal. 
These wavelengths are  490, 500, and 510 nm.
Observe the transmittance profiles with wavelengths of 490 nm and 500 nm can be viewed as a downwards location shift of the transmittance profile at 550nm. 
The same cannot be said for the transmittance profile at 510 nm as the transmittance between inspection points 10 and 15 are nearly identical between the two profiles. 
This indicates both a low signal-to-noise ratio and a functional difference other than a location shift. 
Figure \ref{fig:lowE} also shows how in addition to differences in the mean there are changes in correlation structure. 
Despite the transmittance profile at 510nm having a mean structure very similar to the profile at 550nm, we can see there is a difference in the within-profile correlation structure. 
In simulating the online performance of our control chart, we randomly permute the order of the in-control (or out-of-control) profiles and apply our control chart on the reordered profiles until an alarm is raised. 

Figure \ref{fig:ARL1_FAR_lowE} shows the performance of our control chart using two aggregation methods. 
We observe in nearly all cases the out-of-control profile is detected quickly, but the false alarm rates can be fairly large in some instances. 
We first notice the aggregation method using the minimum of the conditional p-values has a rather large FAR. 
As this may due to the relatively small value of $m$ as compared to what was used in the simulation studies, we combine two adjacent wavelengths 550nm and 545nm to define the set of in-control profiles to artificially increase the number of historical in-control profiles available. 
Observe the FAR does decrease after this artificial increase in $m$. 

The method of aggregating the conditional p-values using a geometric mean results in much lower FARs when compared to the other method of aggregation. 
We additionally see a much more drastic reduction in FAR when the set of in-control profiles are artificially increased in size.

\section{Discussion}

We have proposed a method for online monitoring of random functions which exhibit two forms of randomness and variation, which include randomness in the underlying functional through a Gaussian process applied to the coefficients of a basis expansion of the function and through an additive independent noise term. A key element of this work lies in exploiting the correlation structure of the observed response vectors, which allowed us to detect deviations from in-control behavior that would potentially be missed by alternative methodology developed to primarily monitor valued based deviations (e.g., mean shifts). What is more, the proposed methodology is based on response modeling, as opposed to functional estimation. 
As a result, we do not require a significant number of observation points to estimate each function, reducing the data collection burden as well as providing a scalable methodology. 

While we have demonstrated the potential of this methodology, there are important directions for future research. First, we have assumed a sufficient historical data set from which to learn the in-control process. Second, the number of parameters to estimate scales with the number of monitor sites, which presents a potential challenge in including additional monitor sites if the historical data is not similarly enhanced. Second, the Gaussianity assumption may be too strong for certain applications.  Our proposed monitoring statistic and control chart is inspired by the Markov random field literature, which provides avenues to propose new approaches and methodology to address these challenges in scenarios where they would become relevant. The generality of undirected graphical models (the the probabilistic foundations of Markov random fields) provides a general approach to modeling dependence and interactions which can be exploited to extend the presented methodology beyond Gaussian data. Functional data provides an opportunity to reduce the number of parameters through both parsimonious dependence structure focusing on localized dependence structured around the functional domain, as well as endowing models with additional regularity assumptions that facilitate learning many parameters of interest when the available historical data may be limited. The developments in this work would naturally extend into and benefit from both of these directions which would enable the methodology to be even more widely applicable. 
Lastly, the use of conditional p-values lends itself to developing procedures for identifying out-of-control monitors, which we leave for future work.

\section*{Acknowledgements}
We thank Chancellor Johnstone for communicating Lemma 1 in the supplementary materials and the corresponding proof to us.

%\bibliography{article}

\end{document}

\bibliographystyle{plainnat}

\def\spacingset#1{\renewcommand{\baselinestretch}%
{#1}\small\normalsize} \spacingset{1}

%%%%%%%%%%%%%%%%%%%%%%%%%%%%%%%%%%%%%%%%%%%%%%%%%%%%%%%%%%%%%%%%%%%%%%%%%%%%%%

\if1\blind
{
  \title{\bf Supplementary Materials: Profile monitoring of random functions with Gaussian process basis expansions } 
  \author{ 
    Takayuki Iguchi\thanks{This research was supported by the Test Resource Management Center (TRMC) within the Office of the Secretary of Defense (OSD), contract \#FA807518D0002
    and the Environmental Security Technology Certification Program (EW24-7973).
    The views expressed in this article are those of the authors and do not reflect the official policy or position of the United States Air Force, Department of Defense, or the U.S. Government.} \\
    Department of Mathematics \& Statistics, Air Force Institute of Technology\\\\ 
     Jonathan R. Stewart\\
    Department of Statistics, Florida State University \\\\ 
    Eric Chicken \\
    Department of Statistics, Florida State University \\\\
  }
  \maketitle
} \fi

\if0\blind
{
  \bigskip
  \bigskip
  \bigskip
  \begin{center}
    {\LARGE\bf Profile monitoring of random functions with Gaussian process basis expansions} 
\end{center}
  \medskip
} \fi

\bigskip
\begin{abstract}
We consider the problem of online profile monitoring of random functions that admit basis expansions possessing Gaussian coefficients for the purpose of out-of-control state detection. In this work, we consider random functions that feature two sources of variation: additive error and random fluctuations through random coefficients in the basis representation of functions. We focus on a two-phase monitoring problem with a first stage consisting of learning the in-control process and the second stage leveraging the learned process for out-of-control state detection. We outline a learning and monitoring methodology for our problem by exploiting the Gaussian process of the basis coefficients to develop a scalable and effective monitoring methodology for the observed processes that makes weak functional assumptions on the underlying process. We will demonstrate the potential of our method through simulation studies that highlight some of the nuances that emerge in profile monitoring problems with random functions. Lastly, we will discuss potential refinement and extensions of our methodology.
\end{abstract}

\noindent%
{\it Keywords:} statistical process control, profile monitoring, functional data
\vfill

\newpage
\spacingset{1.8} % DON'T change the spacing!

\section{Outline of proofs}

We now provide the lemmas and an outline of the proof for the theorems in the main manuscript which we include below for completeness. 

\begin{theorem}\label{thm: main}
Let $U_1, \dots, U_m$ and $V_1, V_2, \dots$ be independent and identically distributed continuous random variables (i.e., monitoring statistics).
If $W = \min \{t \in \mathbb{Z}^+ : V_t < U_{(k)}\}$ and $2 \leq k < m$,
then $\mbE[W] = \frac{m}{k-1}$.
\end{theorem}
\begin{theorem}\label{thm: convergegeo}
    Assume the setting of Theorem \ref{thm: main}. 
    Let $R = \min \{t \in \mathbb{Z}^+ : V_t < U_{(k^*)}\}$ 
    be a random variable 
    with $k^* = 1 + \frac{m}{\text{ARL}_0}$. 
    Then $R\overset{d}{\rightarrow} G$ as $m \to \infty$, where $G \sim \operatorname{Geometric}\left(\frac{1}{\text{ARL}_0}\right)$.
\end{theorem}

We begin with a previously published fact concerning order statistics, commonly used in conformal prediction \citep{angelopoulos2022gentle, angelopoulos2023conformal}. 

\begin{lemma}\label{lemma: conformal pred trick}
    If continuous random variables $U_1, \dots, U_m, V$ are a random sample, then $\Pr(V <U_{(k)} ) = \frac{k}{m+1}$, where $U_{(k)}$ is the $k^{\text{th}}$ order statistic of $U_1, \dots, U_m$. 
\end{lemma} 

We extend this idea to the setting where the sequence $V_1, V_2, \dots$ are independent and identically distributed with $U_1, \dots, U_m$. 
We now contextualize $U_1, \dots, U_m$ as $m$ monitoring statistics obtained from $m$ historical in-control profiles separately and $V_1, V_2, \dots$ correspond to monitoring statistics during online monitoring while the process is in-control. 
In this light, we can consider 
$U_{(k)}$ to be a lower control limit.
The next few results 
characterize
the probability of a false alarm in this context.

\begin{lemma}\label{lemma: marginal} {\bf No alarms by time $T$.}
    Let $U_1, \dots, U_m$ and $V_1, V_2, \dots$ be independent and identically distributed continuous random variables.
    If $M_T = \min(V_1, \dots, V_T)$,
    then $\Pr(M_T > U_{(k)})  = \binom{m}{k} /\binom{T+m}{k}$.
\end{lemma}
\begin{lemma}\label{lemma: joint} {\bf Probability of False Alarm at time $T>1$.}
    Under the same setting as Lemma \ref{lemma: marginal}, the probability of a false alarm at time $T>1$ is
    $\Pr\left( V_T < U_{(k)} , \, M_{T-1} \geq U_{(k)}\right) =  \binom{m}{k}
    \frac{k}{(T+m)\binom{T+m-1}{k}}$.
\end{lemma}

Notice setting the lower control limit as $U_{(k)}$ implies the event of a false alarm at time $t$ and the event of a false alarm at time $t+1$ are dependent. 
For example, 
$$\Pr\left( V_2 < U_{(k)} , \, V_1 \geq U_{(k)}\right)
= \frac{k}{m+1}\frac{m-k+1}{m+2}
\neq \frac{k}{m+1}\frac{m-k+1}{m+1}
= \Pr(V_2 < U_{(k)})\Pr(V_1 \geq U_{(k)}).
$$
Since the event of a false alarm at differing points in time are dependent, the in-control run length cannot be modeled as a Geometric random variable despite the fact that the monitoring statistics  at different points in time are in fact independent. 
The assumption that run lengths being geometric is critical for calibrating a control limit to achieve a desired $\text{ARL}_0$ for many control charts in the literature. 
Using Lemma \ref{lemma: joint}, we can obtain an expression for the $\text{ARL}_0$:
\begin{align}\label{eq: ARL0 complicated}
    \text{ARL}_0 
    & = \sum_{t=1}^\infty t \, \Pr(\text{False alarm at time }t)
     = \frac{k}{m+1} + \sum_{t=2}^\infty t \,\, \binom{m}{k}
    \frac{k}{(t+m)\binom{t-1+m}{k}}.
\end{align}
Although this expression can be approximated by truncating the series, our result in Theorem \ref{thm: main} provides a much simpler expression for $\text{ARL}_0$, which has the advantage of being easier to implement in practice. 

\section{Proofs}

\subsection*{Proof of Lemma \ref{lemma: conformal pred trick}}
Let $F$ and $f$ be the CDF and PDF of the shared distribution of $U_1, \dots, U_m, V$,
respectively. 
Due to a standard result \citep[eq. 2.1.6]{arts2004NonparametricPredictiveCC} regarding the density of the $k^{\text{th}}$ order statistic, 
we obtain the joint density of $(U_{(k)}, \, V)$:
$$
f_{U_{(k)}, V} (x,y) 
= f_{U_{(k)}}(x) \; f_V(y)
= k \binom{m}{k} f(x) [F(x)]^{k-1} [1-F(x)]^{m-k} f(y).
$$
We compute the probability $\Pr(V < U_{(k)})$ 
by 
\begin{align*}
\Pr(V < U_{(k)})  
&= \int_{-\infty}^\infty \int_{-\infty}^x  k \binom{m}{k} f(x) [F(x)]^{k-1} [1-F(x)]^{m-k} f(y) \, dy\, dx \\
&= \int_{-\infty}^\infty k \binom{m}{k} f(x) [F(x)]^{k} [1-F(x)]^{m-k}  \, dx.
\end{align*}
identifying the CDF for $V$ as in the inner integral and leveraging the identical distribution assumption 
Letting $u = F(x)$ and pulling out appropriate constants further simplifies to 
$$\Pr(V < U_{(k)})  = k \binom{m}{k} \int_0^1 u^k (1-u)^{m-k} \, du.$$
Recognizing the beta kernel, 
the integral is equal to $\frac{m!}{(k-1)! (m-k)!} \frac{k! (m-k)!}{(m+1)!}$ which further simplifies to $\frac{k}{m+1}$.
\hfill$\square$

\subsection*{Proof of Lemma \ref{lemma: marginal}}
Consider the probability 
$\Pr(M_T \geq U_{(k)})$,
which is the probability 
of no false alarms by time $T$.
By the independence of the random variables $U_1, \ldots, U_m, V$, 
the joint density $ f_{U_{(k)}, M_T}(x,y)$ of $(U_{(k)}, M_T)$  is the product of marginal densities 
$f_{U_{(k)}}(x)$ and $f_{M_T}(y)$, 
allowing us to write: 
$$\Pr(M_T \geq U_{(k)}) = \int_{-\infty}^\infty \int_x^\infty f_{U_{(k)}, M_T}(x,y) \, dy \, dx = \int_{-\infty}^\infty f_{U_{(k)}}(x) \int_x^\infty f_{M_T}(y) \, dy \, dx.$$
Noting that $M_T = \min(V_1, \ldots, V_T)$ 
and using the the density of the minimum of a random sample,
the inner integral can be written as $$ \int_x^\infty f_{M_T}(y) \, dy 
= \int_x^\infty T f(y) [1-F(y)]^{t-1} \, dy.$$
By a change of variable taking $u = 1- F(y)$, 
we can write 
$$\int_x^\infty T f(y) [1-F(y)]^{t-1} \, dy = \int_{0}^{1-F(x)}  T u^{t-1} \, du = [1-F(x)]^{T},$$
showing 
$$\Pr(M_T \geq U_{(k)}) = \int_{-\infty}^\infty f_{U_{(k)}}(x) [1-F(x)]^{T}  \, dx.$$ 
Writing out the density for the $k^{\text{th}}$ order statistic $U_{(k)}$ based on the random sample $U_1, \dots, U_m$ and combining like terms gives 
$$\Pr(M_T \geq U_{(k)}) = \int_{-\infty}^\infty k \binom{m}{k} f_{U_{(k)}}(x) [F(x)]^{k-1} [1-F(x)]^{T+m-k}  \, dx.$$
By a change of variables taking $v = F(x)$, 
$$ 
\int_{-\infty}^\infty k \binom{m}{k} f_{U_{(k)}}(x) [F(x)]^{k-1} [1-F(x)]^{T+m-k}  \, dx
= k \binom{m}{k} \int_{-\infty}^\infty  v^{k-1} (1-v)^{T+m-k}  \, dv.$$
Recognizing the Beta kernel, 
the integral can be evaluated as 
$$ \Pr(M_T \geq U_{(k)}) 
= k \binom{m}{k} \int_{-\infty}^\infty  v^{k-1} (1-v)^{T+m-k}  \, dv
=\frac{k \binom{m}{k}}{ (T+m)  \binom{T+m-1}{k-1}}.$$
For positive integers $a,b$, $\binom{a-1}{b-1} = \frac{b}{a}\binom{a}{b}$,
implying that $(T+m)  \binom{T+m-1}{k-1} = k\binom{T+m}{k}$,
establishing 
\[
\pushQED{\qed} 
    \Pr(M_T \geq U_{(k)}) = \frac{\binom{m}{k}}{\binom{T+m}{k}}.  \qedhere
    \popQED
\]
% \end{align*}
% \begin{align*}
%     \Pr(M_T \geq U_{(k)})
%     & = \int_{-\infty}^\infty \int_x^\infty f_{U_{(k)}, M_T}(x,y) \, dy \, dx\\
%     & = \int_{-\infty}^\infty f_{U_{(k)}} (x) \int_x^\infty f_{M_T}(y) \, dy \, dx\\
%     & = \int_{-\infty}^\infty f_{U_{(k)}} (x) \int_x^\infty T f(y) [1-F(y)]^{t-1} \, dy \, dx\\
%     & = \int_{-\infty}^\infty f_{U_{(k)}} (x) \int_{0}^{1-F(x)}  T u^{t-1} \, du \, dx\\
%     & = \int_{-\infty}^\infty f_{U_{(k)}} (x) [1-F(x)]^{T}  \, dx\\
%     & = \int_{-\infty}^\infty k \binom{m}{k} f_(x) [F(x)]^{k-1} [1-F(x)]^{T+m-k}  \, dx\\
%     & = k \binom{m}{k} \int_{-\infty}^\infty  v^{k-1} (1-v)^{T+m-k}  \, dv\\
%     % & = \frac{m!}{(k-1)!(m-k)!} \frac{(k-1)! (t+m-k)!}{(t+m-1)!}
%     & = \frac{k \binom{m}{k}}{ (T+m)  \binom{T+m-1}{k-1}}\\
%     & = \frac{k \binom{m}{k}}{
%     (t+m) \frac{k}{T+m} \binom{T+m}{k}
%     } \\
%     & = \frac{\binom{m}{k}}{\binom{T+m}{k}}
% \end{align*}
% \end{proof}
% \hfill $\square$

\subsection*{Proof of Lemma \ref{lemma: joint}}
% \begin{proof}
The event of a first alarm occurring at time $T$ 
can be expressed as the event 
$\{V_T < U_{(k)}\} \cap \{ M_{T-1} \geq U_{(k)}\}$.
%and $ M_{T-1} \geq U_{(k)}$. 
The joint probability $\Pr\left( V_T < U_{(k)} ,  \, M_{T-1} \geq U_{(k)}   \right)$ can be written as
\begin{align*}
    \Pr\left( V_T < U_{(k)} |  M_{T-1} \geq U_{(k)}   \right)
    \Pr\left(  M_{T-1} \geq U_{(k)}   \right)
    & = 
    \left(1 - \frac{\Pr\left( V_T > U_{(k)} ,  M_{T-1} \geq U_{(k)}  \right)}{\Pr\left(  M_{T-1} \geq U_{(k)}   \right)}\right)
    \Pr\left(  M_{T-1} \geq U_{(k)}   \right)\\
    & = 
    \Pr( M_{T-1} \geq U_{(k)}  ) - \Pr( M_{T} \geq U_{(k)}  ).
\end{align*}
Using Lemma \ref{lemma: marginal}, 
$\Pr( M_{T-1} \geq U_{(k)}  ) - \Pr( M_{T} \geq U_{(k)}  ) 
    =\binom{m}{k} \left(
    \frac{1}{\binom{T-1+m}{k}}
    - 
    \frac{1}{\binom{T+m}{k}}
    \right)
    = \binom{m}{k}
    \frac{\binom{T+m}{k} - \binom{T-1+m}{k}}{\binom{T-1+m}{k}\binom{T+m}{k}}$
.
By Pascal's rule, this probability is $\binom{m}{k}
    \frac{\binom{T-1+m}{k-1}}{\binom{T-1+m}{k}\binom{T+m}{k}}$. 
For positive integers $a,b$, $\binom{a-1}{b-1} = \frac{b}{a}\binom{a}{b}$. 
With this fact, we can pull out a constant and cancel terms to obtain $\binom{m}{k}
    \frac{\binom{T-1+m}{k-1}}{\binom{T-1+m}{k}\binom{T+m}{k}}
    = \binom{m}{k}
    \frac{k}{T+m}\frac{\binom{T+m}{k}}{\binom{T-1+m}{k}\binom{T+m}{k}}
     = \binom{m}{k}
    \frac{k}{(T+m)\binom{T-1+m}{k}}$.
The factorials in the denominator can be rearranged from $(T+m)\binom{T-1+m}{k}$ to $(T+m-k)\binom{T+m}{k}$, providing the desired result. 
\hfill$\square$

\begin{lemma}\label{lemma: series}
    If $a>b>2$ are positive integers, then for $T\geq 1$
    \begin{align*}
        \sum_{t=0}^T t \frac{(t+a-b)!}{(t+a)!}
        = 
        \frac{1}{(b-2)(b-1)}
        \left(
        \frac{(a-b+1)!}{(a-1)!}
        - 
        ((b-1)T + a) 
        \frac{(T+a-b+1)!}{(T+a)!}
        \right)
    \end{align*}
    and moreover 
    $    \sum_{t=0}^\infty t \frac{(t+a-b)!}{(t+a)!}
        = 
         \frac{(a-b+1)!}{(b-2)(b-1)(a-1)!}
    $.
\end{lemma}
\subsection*{Proof of Lemma \ref{lemma: series}}
    We prove the result regarding the partial sum by way of induction on $T$. 
    We begin with the base case: 
    \begin{align*}
        \frac{\frac{(a-b+1)!}{(a-1)!} - (a+b-1) \frac{(a-b+2)!}{(a+1)!}}{(b-2)(b-1)}
        & =
        \frac{(a-b+1)!}{(b-2)(b-1)(a+1)!}
        \left(a(a+1) - (a+b-1) (a-b+2)\right)\\
        & =
        \frac{(a-b+1)!}{(b-2)(b-1)(a+1)!}
        \left(a^2 + a - a^2 - a + (b-1)(b-2)\right)\\
        & =
        \frac{(a-b+1)!}{(a+1)!}.
    \end{align*}
    For the inductive step, assume $T$ is a value for which the desired result regarding the partial sum is true. 
    \begin{align*}
        \sum_{t=0}^{T+1} t \frac{(t+a-b)!}{(t+a)!}
        & =
        \frac{(a-b+1)!}{(b-2)(b-1)(a-1)!}
        - 
        ((b-1)T + a) 
        \frac{(T+a-b+1)!}{(T+a)!}
        +
        (T+1) \frac{(T+a-b+1)!}{(T+a+1)!}\\
        & =
        \frac{(a-b+1)!}{(b-2)(b-1)(a-1)!}
        -
        \frac{(T+a-b+1)!}{(b-2)(b-1)(T+a+1)!}\\
        & \qquad\qquad \times
        \left(
            (T(b-1)+a)(T+a+1) 
            -
            (T+1)(b-2)(b-1)
        \right)
    \end{align*}
It suffices to show $$(T(b-1)+a)(T+a+1)-(T+1)(b-2)(b-1) = (T+a-b+2)((T+1)(b-1)+a)$$ to complete the inductive step. 
\begin{align*}
    (T(b-1)+a)(T+a+1)\\
    \qquad\qquad\qquad-(T+1)(b-2)(b-1)
    & =
    (b-1)T^2 + ((b-1) (a-b+2) + (a+b-1))T \\
    & \qquad + a^2 + ((b-1) - (b-2)) a + (b-1)(b-2)\\
    & =
    (b-1)T^2 + ((b-1) (a-b+2) + (a+b-1))T \\
    & \qquad + (a + b-1)(a - b + 2)\\
    & = 
    ((b-1)T + a + b - 1) (T+a - b +2)\\
    & =
    ((T+1)(b-1)+a)(T+a-b+2).
\end{align*}
The result regarding the partial sum has been proved. 
To show the result regarding the series, it suffices to show $\lim\limits_{T\to\infty} ((b-1)T + a)\frac{(T+a-b+1)!}{(T+a)!} = 0.$
We apply the upper and lower bounds on a factorial provided by \citet[p. 1]{robbins1955remarkStirling} to obtain
$((b-1)T + a)\frac{(T+a-b+1)!}{(T+a)!} \in \left(c(T) e^{c_0 + \frac{1}{12 T + c_1} - \frac{1}{12 T + c_2}}, c(T) e^{c_0 + \frac{1}{12 T + c_3} - \frac{1}{12 T + c_4}}\right)$ for constants $c_0, c_1, c_2, c_3, c_4$ and where $$c(T) = ((b-1)T + a) \sqrt{\frac{T+a-b+1}{T+a}} \frac{(T + a - b + 1)^{T + a - b + 1}}{(T + a )^{T + a }}.$$
Observing the following limits 
    $\lim\limits_{T\to\infty}\sqrt{\frac{T+a-b+1}{T+a}} = 1$,
    $\lim\limits_{T\to\infty} \left(\frac{T + a - b + 1}{T + a }\right)^{T} = e^{1-b}$,
    $\lim\limits_{T\to\infty}\left(\frac{T + a - b + 1}{T + a }\right)^{a-b+1} = 1$, and
    $\lim\limits_{T\to\infty} \frac{T + \frac{a}{b-1}}{(T+a)^{b-1}} = 0$
yields $\lim\limits_{T\to\infty} C(T) = 0$ and by extension, the desired limit is also zero. 
\hfill$\square$

\begin{corollary}\label{corollary: series}
    Applying $a = m$ and $b=k+1$ with $k\geq 2$ to Lemma \ref{lemma: series} yields
    $
    \sum_{t=0}^{\infty} t \frac{(t+m-k-1)!}{(t+m)!} 
        = 
        \frac{(m-k)!}{k(k-1)(m-1)!}
    $.
\end{corollary}

\subsection*{Proof of Theorem \ref{thm: main}}
    By definition of expectation and Lemma \ref{lemma: joint}, $\mbE[W]$ is given by \eqref{eq: ARL0 complicated}.
    We add a convenient form of zero to obtain $\mbE[W]
     = \frac{k}{m+1} + k \binom{m}{k} \left( \frac{-1}{(m+1) \binom{m}{k}} + 
        \sum_{t=0}^\infty \frac{t}{(t+m)\binom{t+m-1}{k}}  
    \right)$.
    Cancelling terms and bringing $\binom{m}{k}$ inside the sum yields $
    \mbE[W] = k \sum_{t=2}^\infty \frac{t}{(t+m)}\frac{m!}{k!(m-k)!}\frac{k! (t+m-1-k)!}{(t+m-1)!}$ and after simplifying we obtain $k\frac{m!}{(m-k)!} \sum_{t=0}^\infty t\frac{(t+m-k-1)!}{(t+m)!}$. 
    We apply Corollary \ref{corollary: series} to get $\mbE[W] = k\frac{m!}{(m-k)!} \frac{(m-k)!}{k(k-1)(m-1)!}$ and simplify to obtain $\mbE[W] =\frac{m}{k-1}$. 
    $\hfill \square$

\subsection*{Proof of Theorem \ref{thm: convergegeo}}
It suffices to show that the PMF of $W$ converges point-wise to the PMF of a Geometric distribution. 
By Lemma \ref{lemma: joint}, 
this amounts to showing,
for all $T \in \{0, 1, 2, \ldots\}$,
that 
\begin{equation}
\label{eq:key_limit}
\lim\limits_{m\to \infty} \Pr(W = T)
=\lim\limits_{m\to \infty} \binom{m}{k} \frac{k}{(T+m-k) \binom{T+m}{k}} = \frac{1}{A_0} \left(1-\frac{1}{A_0}\right)^{T-1},
\end{equation}
denoting $\text{ARL}_0$ by $A_0$ 
for ease of presentation. 
Letting $A^\star = 1 - 1 \,/\, A_0$ 
and applying Theorem \ref{thm: main},
we have 
$m-k = A^\star \, m - 1$ and $T + m -k = A^\star \, m + T - 1$.
The PMF of $W$ can then be rewritten as  
$$\Pr(W = T) 
= \frac{B \, (m+ A_0)}{A_0 \, (A^\star \,  m + T - 1)},
\quad\quad  \text{where}\;
B = \frac{m! \, (A^\star \, m + T - 1)!}{(T+m)!\, (A^\star \, m  - 1)!}.$$ 
Define the following functions for $m \in \mathbb{Z}^+$: 
\begin{align*}
    c_1(m) 
    & 
    = \sqrt{\frac{m \, ( A^\star \,  m + T - 1) }{(T+m) \, (A^\star \,  m  - 1)}}\\\\
    c_2(m) 
    & = \frac{m^m \, (A^\star \,  m + T - 1)^{A^\star \,  m + T - 1}}{(T+m)^{T+m} \, (A^\star \,  m - 1)^{A^\star \,  m  - 1} }\\\\
    c_3(m) 
    & =  \exp\left( \frac{-1}{12 \, m+1} - \frac{1}{12 \, (A^\star \,  m + T - 1)+1} + \frac{1}{12\, (A^\star \,  m - 1)} + \frac{1}{12 \, (T+m)} \right)\\\\
    c_4(m) 
    & = \exp\left( \frac{1}{12 \, m} + \frac{1}{12 \, (A^\star \,  m + T - 1)} - \frac{1}{12 \, (A^\star \,  m  - 1)+1} - \frac{1}{12 \, (T+m)+1} \right).
\end{align*}
According to  \citet[p. 1]{robbins1955remarkStirling}, %
we can upper and lower bound $B$ using a variant of Stirling's formula for any value of $m$ using the previously defined functions:  
$$c_1(m) \, c_2(m) \, c_3(m) < B 
= \frac{m! \, (A^\star \, m + T - 1)!}{(T+m)!\, (A^\star \, m  - 1)!}
< c_1(m) \, c_2(m) \, c_4(m).$$
As $\lim\limits_{m\to\infty} [a \, m + b]^{-1} = 0$ for constants $a \in \mathbb{R}$ and $b \in \mathbb{R}$ such that $a \, m + b \neq 0$, 
we have the limits 
$$\lim\limits_{m\to\infty} \, c_3(m) 
= \lim\limits_{m\to \infty} \, c_4(m) = 1.$$
By the squeeze theorem, it therefore suffices to show
that 
\begin{equation} 
\label{eq:need_to_show}
\lim\limits_{m\to\infty} \, 
 \frac{c_1(m) \, c_2(m) \, ( m + A_0)}{A^\star\, m + T - 1}  = (A^\star)^{T-1}
\end{equation}
in order to demonstrate \eqref{eq:key_limit} to establish 
$V\overset{d}{\rightarrow} G$
where $G \sim\operatorname{Geometric}\left(\frac{1}{\text{ARL}_0}\right)$. 

We begin with the limit of $c_1(m)$: 
$$
\lim\limits_{m\to\infty} \, c_1(m)
= \lim\limits_{m\to\infty} \sqrt{\frac{m \, ( A^\star \, m + T - 1) }{(T+m) \, (A^\star \, m  - 1)}} = 
\lim\limits_{m\to\infty} 
\sqrt{\frac{m}{(T+m)}}
\sqrt{\frac{( A^\star \, m + T - 1) }{
(A^\star \, m  - 1)}} = 1.$$
Next, 
we rewrite $c_2(m)$ as 
$$c(m) = \left(\frac{m}{T+m}\right)^m 
\left(\frac{A^\star \, m + T - 1}{A^\star \, m - 1}\right)^{A^\star \, m} \frac{(A^\star \, m - 1)(A^\star \, m + T - 1)^{T-1}}{(m+T)^T}.$$
In general, 
for constants $(a,b,c) \in \mathbb{R}^3$ such that 
$a \, m + c \neq 0$, 
we have  
$$
e^{b-c}
= \lim\limits_{m\to\infty}  
\left( 1 + \dfrac{b-c}{a \, m + c}\right)^{a \, m+ c}
= \lim\limits_{m\to\infty} 
\left(\frac{a \, m + b}{a \, m + c}\right)^{a \, m},$$
which shows the limits 
$$\lim\limits_{m\to\infty} \left(\frac{m}{T+m}\right)^m = e^{-T}
\quad\quad \text{and} \quad\quad 
\lim\limits_{m\to\infty} 
\left(\frac{A^\star \,  m + T - 1}{A^\star\,  m - 1}\right)^{A^\star \, m} = e^{T}.$$
Next, 
we tackle the limit of the remaining unaddressed term in $c_2(m)$:   
$$\lim\limits_{m\to\infty} 
\frac{(A^\star \,  m - 1) \, (A^\star \,  m + T - 1)^{T-1}}{(m+T)^T} = (A^\star)^T,$$
which shows that 
$$\lim\limits_{m\to\infty} 
c_2(m) = e^{-T} \, e^{T} \, (A^*)^T = (A^*)^T.$$
Lastly, 
$$\lim\limits_{m\to\infty} 
\, \frac{m+A_0}{A^\star \,  m + T -1} 
= \dfrac{1}{A^\star}.$$ 
Combining results, 
we return to the limit in \eqref{eq:need_to_show} and write 
$$\lim\limits_{m\to\infty} 
\left(\frac{ m + A_0}{A^\star \,  m + T - 1}\right) \, c_1(m) \, c_2(m)  = \left(\frac{1}{A^\star}\right) \, (1) \, (A^\star)^{T} = (A^\star)^{T-1},$$
which demonstrates the desired result. 
\hfill $\square$

\section{Motivation for isolating OOC monitors}

\begin{figure}
    \centering
    \includegraphics[width=\linewidth]{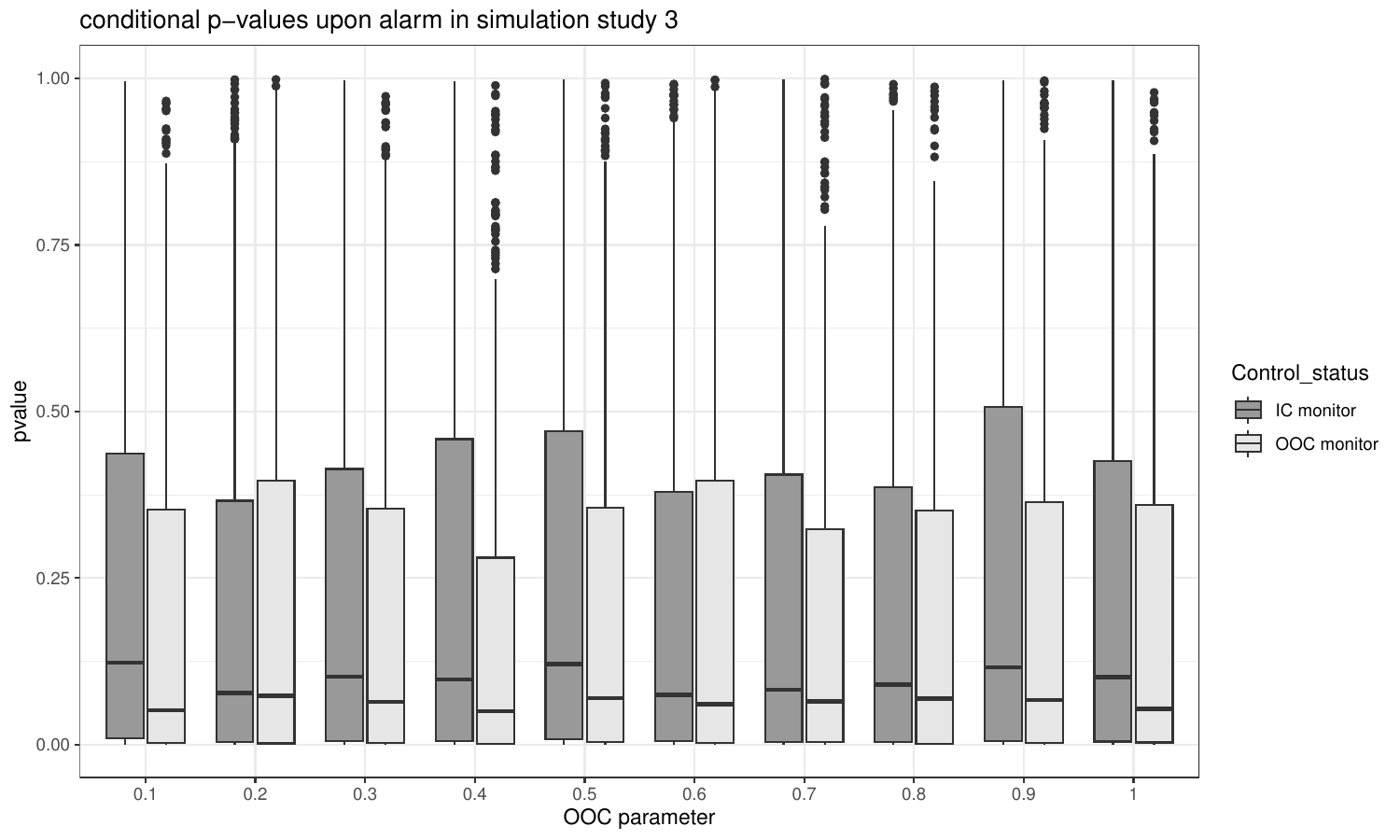}
    \caption{The conditional p-values computed upon alarm in simulation study 3. 
    The conditional p-values are grouped by the true control status of the monitors in simulation study 3.
    }
    \label{fig:condpvals}
\end{figure}

Due the use of conditional p-values, it should be possible to devise a method to isolate the monitors which OOC when our proposed control chart raises an alarm. 
To provide some level of justification for this, Figure \ref{fig:condpvals} shows how the conditional p-values in simulation study 3 tend to be smaller for the OOC monitors than the IC monitors. 
We leave the identification of OOC monitors as future work. 

%\bibliography{article}